\documentclass[aps,twocolumn,floats,prd,nofootinbib,superscriptaddress]
{revtex4-2}

\usepackage{amsmath,amsfonts,amssymb}
\usepackage{graphicx}
\usepackage{dcolumn}
\usepackage{appendix}
\usepackage{subfigure}
\usepackage{placeins}
\usepackage[usenames,dvipsnames]{xcolor} 
\usepackage[export]{adjustbox}
\usepackage{bm}
\usepackage{hyperref}
\usepackage{adjustbox}
\usepackage{tabularx}
\hypersetup{
	colorlinks=true,
	linkcolor=red,
	filecolor=magenta,      
	urlcolor=blue,
}
\usepackage[normalem]{ulem}

\def\Lx{L_{\rm X<2 \, keV}/{\rm SFR}}
\def\Lu{\, {\rm erg} \;{\rm s}^{-1} \;{\rm M}_{\odot}^{-1} \;{\rm yr}}

\newcommand{\RNum}[1]{\uppercase\expandafter{\romannumeral #1\relax}}

\begin{document}

\preprint{}

\title{HERA Bound on x-ray luminosity when accounting for population III stars}
\author{Hovav Lazare}
\email{hovavl@post.bgu.ac.il}
\affiliation{Department of Physics, Ben-Gurion University of the Negev, Be’er Sheva 84105, Israel}

\author{Debanjan Sarkar}
 \email{debanjan@post.bgu.ac.il}
\affiliation{Department of Physics, Ben-Gurion University of the Negev, Be’er Sheva 84105, Israel}

\author{Ely D. Kovetz}
\email{kovetz@bgu.ac.il}
\affiliation{Department of Physics, Ben-Gurion University of the Negev, Be’er Sheva 84105, Israel}


\begin{abstract}
Recent upper bounds from the Hydrogen Epoch of Reionization Array (HERA) on the cosmological 21-cm power spectrum at redshifts $z \approx 8, 10$, have been used to constrain $\Lx$, the soft-band X-ray luminosity measured per unit star formation rate (SFR),  strongly disfavoring values lower than $\approx 10^{39.5} \Lu$. 
This conclusion is derived from semi-numerical models of the 21-cm signal, specifically focusing on contributions from atomic cooling galaxies that host PopII stars.
In this work, we first reproduce the bounds on $\Lx$ and other parameters using a pipeline that combines machine learning emulators for the power spectra and the intergalactic medium characteristics, together with a standard Markov chain Monte Carlo parameter fit. We then use this approach when including molecular cooling galaxies that host PopIII stars in the cosmic dawn 21-cm signal, and show that lower values of $\Lx$ are hence no longer strongly disfavored. 
The revised HERA bound does not require high-redshift X-ray sources to be significantly more luminous than high-mass X-ray binaries observed at low redshift.

\end{abstract}

\maketitle


\section{Introduction}
\label{sec:intro}

The 21-cm signal carries immense potential to probe the cosmic dawn and epoch of reionization, two pivotal phases in the early Universe \cite{Madau:1996cs, Loeb:2003ya, Pritchard:2011xb, Pober:2013jna, Liu:2022iyy}. 
It arises from the hyperfine transition of neutral hydrogen atoms, and its measurement enables us to peer into the Universe's infancy \cite{Furlanetto:2006jb}. 
By studying the fluctuations in the 21-cm signal, we can unravel the processes that drove the formation of 
the first stars and galaxies \cite{Barkana:2000fd, Barkana:2004vb, Chen:2006zr}, 
as well as the subsequent ionization of the intergalactic medium (IGM) \cite{Miralda-Escude:1998adl, Furlanetto:2004nh}. 
The 21-cm signal is a powerful tool for understanding cosmic evolution and 
unlocking the mysteries of our cosmic origins \cite{Loeb:2003ya, Pritchard:2011xb, Chen:2016zuu}.

Several groundbreaking experiments have been designed to target the 21-cm signal~
\cite{Mellema:2012ht, Beardsley:2016njr, Bowman:2018yin, Trott:2019lap, Greig:2020suk, Mertens:2020llj, Ansari:2022nmy},  
including the  Hydrogen Epoch of Reionization Array (HERA)\footnote{https://reionization.org/}
\cite{DeBoer:2016tnn}. 
HERA has recently released two data sets \cite{HERA:2021bsv,HERA:2022wmy}, that were used to set an upper bound on the 
21-cm fluctuation power spectrum from the epoch of reionization ($z\!\sim\!8$ and $10$). 
This upper limit \cite{HERA:2021bsv, HERA:2022wmy}, 
together with other observations, including the galaxy ultraviolet (UV) 
luminosity function \cite{Bouwens:2014fua,Bouwens:2016aaa,Oesch:2017abc,Gillet:2019fjd}, 
the reionization optical depth \cite{Planck:2018vyg} and 
the IGM neutral fraction \cite{McGreer:2014qwa} was then used to place the first constraints on properties related 
to the evolution of the first galaxies and the epoch of reionization
~\cite{HERA:2021noe}.

A major finding of Ref.~\cite{HERA:2021noe} has to do with the parameter $L_{\rm X<2 \, keV}/{\rm SFR}$ that quantifies the X-ray luminosity 
sourced by the endpoints of the stars that drive cosmic dawn and is defined as the ratio of the integrated soft-band ($<2$ keV) 
X-ray luminosity to the star formation rate (SFR).
The HERA upper limit \cite{HERA:2021bsv, HERA:2022wmy} disfavors
$\Lx \lesssim 10^{40}\, {\rm erg} \;{\rm s}^{-1} \;{\rm M}_{\odot}^{-1} \;{\rm yr}$ 
values with high significance. 
Taken at face value, this  would imply that  galaxies at high redshifts
were more X-ray luminous than metal-enriched galaxies in the local Universe~\cite{Mineo:2011id,Lehmer:2010en}.

\begin{figure}[h!]
    \centering
    \includegraphics[width = \columnwidth]{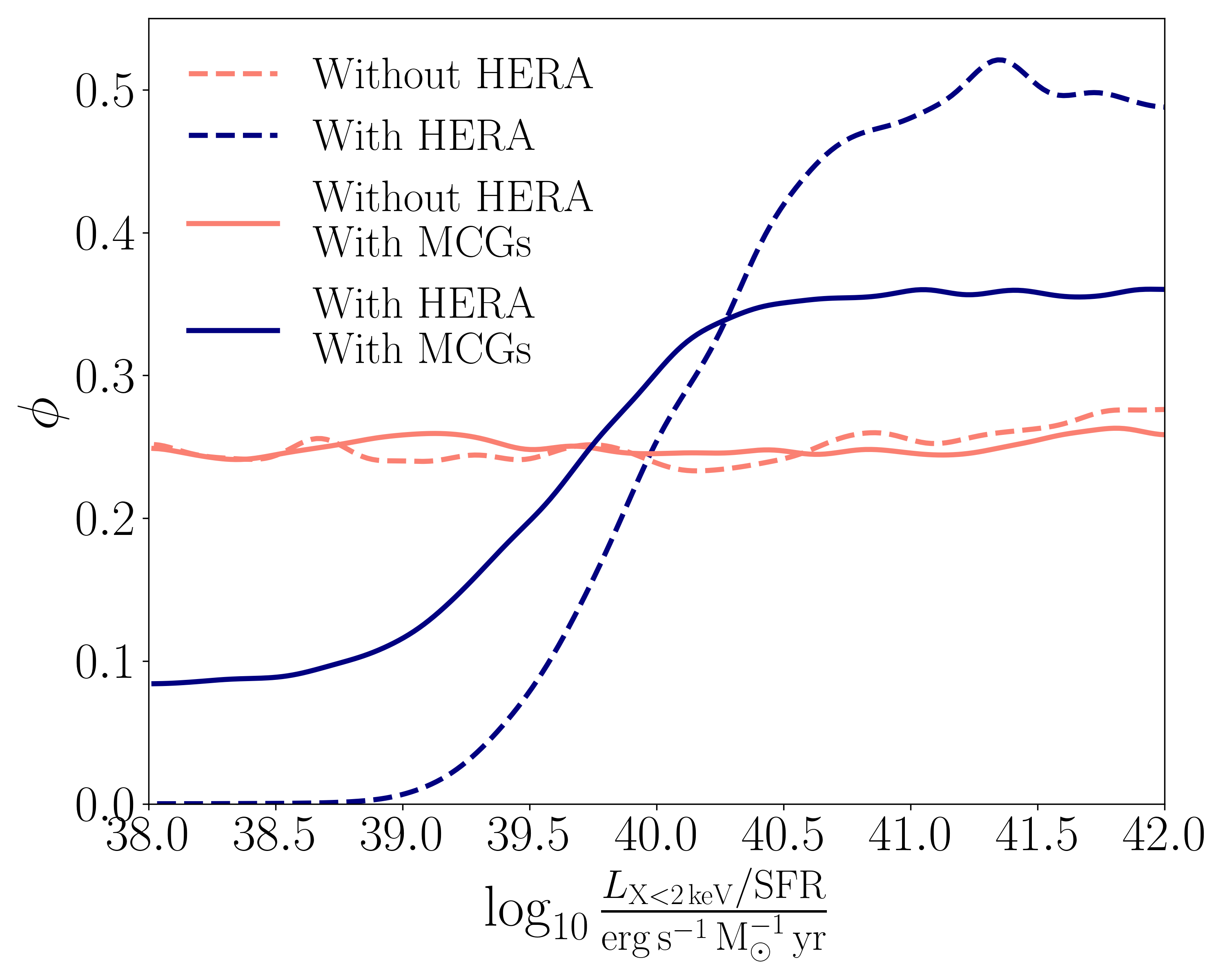}
    \vspace{-0.35in}
    \caption{Marginalized posteriors for $L_{\rm X<2 \, keV}/{\rm SFR}$, with ({\it blue} and without ({\it orange}) HERA and with ({\it solid}) and without ({\it dashed}) molecular cooling halos (MCGs). With the inclusion of MCGs, the preference for strong X-ray heating weakens.}
    \label{fig:Lx_marg_all}
\end{figure}

Crucially, the analysis in Ref.~\cite{HERA:2021noe} considered only  
Population II (or PopII) stars~\cite{Smith:2015vpa} as the sources of heating and ionization.
The first generation of stars (PopIII)~\cite{Bromm:2003vv, Bromm:2005gs} 
are believed to have formed inside metal-poor galaxies 
(sometimes called molecular cooling galaxies or MCGs) hosted by mini halos where gas is cooled by 
molecular line transitions \cite{Haiman:1995jy, Haiman:1996rc, Tegmark:1996yt, Abel:2001pr, Bromm:2003vv, 
Yoshida:2006bz, Haiman:2006si, Trenti:2010hs}. 
While we have
very limited knowledge about the formation and evolution of MCGs,
a number of simulations that have modelled them have shown that they can significantly prepone the onset of cosmic dawn~
\cite{Kulkarni:2020ovu, Mebane:2020jwl, Mirocha:2017xxz, Qin:2020xyh, Qin:2020pdx, Munoz:2021psm}. 
In addition, MCGs  contribute to the total X-ray and ionizing photon 
budget of the Universe \cite{Xu:2014aja, Xu:2016tkk}. 
Ignoring MCGs in any analysis could therefore lead to an 
overestimation of the X-ray and ionizing efficiencies of the atomic cooling galaxies (ACGs)
that are believed to host mostly PopII stars \cite{Qin:2020pdx}. 
This concern has also been raised in Ref.~\cite{HERA:2022wmy}.

In this paper, we reanalyze the HERA data  including MCGs, using machine learning to emulate the 21-cm signal~\cite{Kern:2017ccn, Jennings:2018eko, Cohen:2019vck, Bye:2021ngm, Bevins:2021eah, Choudhury:2021ybn,Sikder:2022hzk} and other global quantities. The implication for the $\Lx$ constraint is  summarized in Fig.~1.


 Simulations play a crucial role in modelling the 21cm signal and interpreting the observational data~
 \cite{Ciardi:2000wv, Santos:2009zk, Shapiro:2012xha, Hassan:2015aba, Sarkar:2016lvb, Ghara:2017vby, Mesinger:2017paq, 
 Sarkar:2018gcb, Ocvirk:2018pqh, Majumdar:2014cza, Sarkar:2019nak, Molaro:2019mew, Kannan:2021xoz, Shaw:2022fre, Munoz:2023kkg, Maity:2022usv, Munoz:2019fkt, Munoz:2019rhi}. 
 Due to the complexity of the astrophysical processes involved, simulations help us understand the 
 underlying physics and generate theoretical predictions for the 21-cm signal. However, simulations 
 can be computationally expensive, if we need to simulate these complex processes accurately. 
 Furthermore, in the traditional Bayesian parameter-inference pipeline, where we use simulations in each
 step to compare the outcomes against observations, we require a significant amount of time before the
 process converges. Even with faster semi-numerical codes such as  \texttt{21cmFAST}\footnote{\href{https://github.com/21cmfast/21cmFAST}{github.com/21cmfast/21cmFAST}}~\cite{Mesinger:2010ne},
 the MCMC process with a minimal number of parameters can take up to several weeks to converge. 
 We, therefore, need alternative methods.

 Machine learning (ML) techniques have emerged as powerful tools for emulating the 21-cm signal and circumventing the time-consuming nature of traditional simulators~\cite{Kern:2017ccn, Jennings:2018eko, Cohen:2019vck, Bye:2021ngm, Bevins:2021eah, Choudhury:2021ybn,Sikder:2022hzk}. The importance of machine 
 learning in this context lies in its ability to significantly accelerate the parameter inference
 process. By training a machine learning model on a large set of pre-computed simulations, 
 it becomes possible to generate fast and accurate emulators that can predict the 21-cm signal 
 for a given set of cosmological and astrophysical parameters. 
 Moreover, machine learning emulators can be trained 
 on a diverse range of simulations, enabling the exploration of different cosmological 
 scenarios and astrophysical processes. 

With the goal of revisiting the analysis in 
Ref.~\cite{HERA:2021noe} to check the robustness of the  $\Lx$ constraint to the inclusion of the PopIII stars or MCGs contribution, we have built an ML-based emulator, that
emulates the 21-cm power spectrum $\Delta^2_{21}(k)$ as a function of wave vector $k$ at different
redshifts ranging from cosmic dawn to the epoch of reionization. We also built separate emulators for 
 the CMB optical depth and IGM neutral fraction.

Our paper is organized as follows. 
In Section~\ref{section_new} we briefly describe the models used in the HERA analysis~\cite{HERA:2021noe} and the results of their analysis.
In Section~\ref{section_2}
we describe the HERA phase-I data and the likelihood we use for our analysis, including the additional observables. In Section~\ref{section_3} we describe the artificial neural network architecture our emulator is based on, how we build our training, validation and test sets, and demonstrate the performance of emulator. 
In Section~\ref{sec:role_of_mhs} we explain how we include the PopIII contribution in the modeling of the 21-cm signal. We then present our results in Section~\ref{section_5}.
After reproducing the HERA parameter constraints with our pipeline when considering only PopII stars, we include the  MCGs in our simulations and
train corresponding emulators.
We then perform the MCMC analysis by considering both the ACG and MCG parameters.
In particular, we present the marginalized posterior of $\Lx$ from this
analysis (summarized in Fig.~\ref{fig:Lx_marg_all}). We find that with the mini halos, we can no longer discard the 
$\Lx \lesssim 10^{40}\, {\rm erg} \;{\rm s}^{-1} \;{\rm M}_{\odot}^{-1} \;{\rm yr}$ 
X-ray luminosity values easily. Although there is a small decline
in posterior probability at 
$\Lx < 10^{39.5} \, {\rm erg} \;{\rm s}^{-1} \;{\rm M}_{\odot}^{-1} \;{\rm yr}$, 
the $L_{\rm X}/{\rm SFR} \sim 10^{39} \, {\rm erg} \;{\rm s}^{-1} \;{\rm M}_{\odot}^{-1} \;{\rm yr}$ 
values are still allowed. This suggests that the relationship between star formation and
soft X-ray luminosity for high redshift galaxies may not be very different 
from the local galaxies \cite{Lehmer:2010en, Mineo:2011id}. 
We discuss and elaborate on this conclusion in Section~\ref{section_6}.

\section{Summary of the HERA Analysis}
\label{section_new}

Ref.~\cite{HERA:2021noe} have considered four different models of reionization to interpret the HERA 
observation. We briefly discuss the models below, and the inferences based on those. 

 \begin{itemize}
     \item {\bf Density-driven linear bias model:}
     In this model, the 21-cm fluctuations are assumed to follow the density fluctuations and the 21-cm
     power spectrum is proportional to the matter power spectrum. A bias parameter, which is a function of quantities like $x_{\rm HI}$, $T_{\rm S}$ and $T_{\rm CMB}$, relates the two power spectra. 

     \item {\bf Phenomenological reionization-driven model:} 
     In this model, IGM is considered as a two phased 
     system consisting of fully ionized bubbles and medium with uniform temperature outside of the 
     bubbles \cite{Mirocha:2022pys}. 

     \item {\bf Semi-numerical model I (\texttt{21cmFAST}):}
     This model assumes that the star-forming galaxies reside inside dark matter haloes and these galaxies
     are responsible for the heating and ionization of the IGM. This utilizes empirical scaling relations 
     to relate galaxy properties to their host dark matter haloes. The galaxy properties include stellar to halo 
     mass ratio, escape fraction of ionizing UV radiation, the star formation rate and X-ray luminosity~\cite{Mesinger:2010ne}.

     \item {\bf Semi-numerical model II:}
     This is an alternate semi-numerical model~\cite{Reis:2021nqf}, 
     built on similar assumptions as \texttt{21cmFAST}. 
     On top of that, this model allows for an excess radio background. 
     Excess radio background, on top of CMB, 
     produces an enhanced 21-cm absorption signal and can also increase the amplitude of 
     the 21-cm power spectrum. 

 \end{itemize}

The first two models relate the 21-cm signal directly to the IGM properties, without trying to model the 
source properties. On the other hand, the semi-numerical models try to model the properties of the 
radiation sources and their evolution in redshift. 

The analysis of Ref.~\cite{HERA:2021noe} based on all of the above models suggest that the IGM at $z=10.4$ 
is heated over the adiabatic cooling limit at $>95\%$ confidence, and this rule out 
the ``cold reionization'' scenarios in which the IGM continues to adiabatically cool until it reionizes.

Considering the Semi-numerical model I, 
Ref.~\cite{HERA:2021noe} quotes a lower limit on the parameter $\Lx$ 
which describes the heating ability of the EoR galaxies per unit of star formation.
We discuss the significance of $\Lx$ later in this paper. 


HERA observations, Based on the Semi-numerical model II, and combined with the Chandra X-ray background constraints, 
rule out most of the models that explain the radio background observed by 
LWA \cite{Dowell:2018mdb} and ARCADE-2 \cite{Fixsen:2009xn} as originating at 
$z\gtrsim8$.

Note that the inferences in Ref.~\cite{HERA:2021noe} are very much dependent on the choice of physical models
and the interpretations of different quantities change with models. 
In our current analysis, we mainly focus on Semi-numerical model I (\texttt{21cmFAST}), 
and we discuss the findings based on this model in detail.

\section{Data and Likelihood}\label{section_2}

\subsection{The 21-cm Observables}

The emission or absorption of the 21-cm line is characterized by the spin temperature, $T_S$, 
and is  usually measured as the differential brightness temperature, $\delta T_{21}$, 
with respect to the brightness temperature of the low-frequency radio background, $T_{\rm{rad}}$.
In the usual scenario, $T_{\rm{rad}}$ is taken to be the CMB temperature $T_{\rm{cmb}}=2.7254 \times (1+z)$ K
where $z$ refers to the cosmological redshift.
The differential brightness temperature, $\delta T_{21}$, can be expressed as~
\cite{Bharadwaj:2004nr, Furlanetto:2009iv, Madau:2015cga} 
\begin{equation}\label{eq:21cm_signal}
    \begin{aligned}
        \delta T_{21}(\nu) &= \frac{T_{\mathrm{S}}-T_{\rm rad}}{1+z}\left(1-e^{-\tau_{\nu_{0}}}\right) \,.
    \end{aligned}
\end{equation}
Here, the factor $\left(1-e^{-\tau_{\nu_{0}}}\right)$ exhibits the effect of propagation 
through a medium, where $\tau_{\nu_{0}}$ is the optical depth. Note that, in this equation, the astrophysical 
information is mostly encoded in $T_{\mathrm{S}}$, whereas the state of the propagating medium 
is captured by the optical depth $\tau_{\nu_{0}}$.

The measurement of  $\delta T_{21}$ is done in two different ways:\\
(i) \textbf{Global signal} : Here, $\delta T_{21}(\mathbf{n},z)$ is first measured along  different directions
$\mathbf{n}$ of the sky, keeping the redshift (or frequency) fixed. The measurements are finally averaged over the 
directions to convert into an average quantity $\langle \delta T_{21}(z) \rangle$, which is known as the 
global signal~\cite{Monsalve:2017mli, Bevins:2022ajf}. 
\\
(ii) \textbf{Fluctuations}: Here, telescopes measure the spatial fluctuations of the 
$\delta T_{21}(\mathbf{x},z)$ field. The fluctuations are then interpreted in terms of  
statistics like the power spectrum
~\cite{Vrbanec:2020zee, Karagiannis:2020dpq, Rahimi:2021wom, Mirocha:2022pys, Cook:2022vlj, 
Berti:2022ilk, Lewis:2023jcz}, 
bispectrum~\cite{Yoshiura:2014ria, Shimabukuro:2015iqa, Sarkar:2019ojl, Durrer:2020orn, 
Cunnington:2021czb, Kamran:2021knv, Karagiannis:2022ylq}, etc. 
The 21-cm  power spectrum is the most relevant quantity for this work, and is defined as 
\begin{equation}
\label{eq:power_spectra}
    \langle \tilde{\delta T}_{21}({\bf k}_1) \tilde{\delta T}_{21}({\bf k}_2) \rangle = (2 \pi)^3 \delta^D({\bf k}_1 - {\bf k}_2) P_{21}({\bf k}_1)\,,
\end{equation}
where $\langle...\rangle$ denotes an ensemble average, $\tilde{\delta T}_{21}({\bf k})$ is the  Fourier transform of $ \delta T_{21}(\bf x)$ and $\delta^D$ is the Dirac delta function. Specifically, the quantity of interest for us is $\Delta_{21}^2({\bf k}) \equiv k^3 P_{21}({\bf k})/(2 \pi^2)$, which will be the output of HERA measurements, and of our simulations and emulators.

\subsection{HERA Phase-I data}

The upper limits on the 21-cm signal from HERA were first published in Ref.~\cite{HERA:2021bsv}, and improved limits were published in Ref. ~\cite{HERA:2022wmy}
These limits are based on 94 nights of observation using the HERA Phase I experimental configuration.
For detailed information on the configuration, the reader is referred to Ref.~\cite{HERA:2022wmy}.
Out of all the different bands of observation, in this work we concentrate on two bands:
Band 2, centered at $z = 7.9$; and Band
1, centered at $z = 10.4$. These bands are
largely free of radio frequency interference (RFI).
Following Ref.~\cite{HERA:2022wmy}, we use the observed power spectrum computed using all fields for each redshift band. 
For a comprehensive reading about the data, the reader is referred to Ref.~\cite{HERA:2022wmy,HERA:2021bsv}.

\subsection{The data Likelihood} 

The form of the likelihood function is important for any Bayesian analysis. The 21-cm power spectrum measurements 
from HERA suffer from (largely unknown) systematics which need to be taken into account in the 
likelihood analysis. Below, we briefly explain how to derive the likelihood in the presence of
systematics. 

First, we divide our data into two vectors, one for every redshift bin,
\begin{equation}\label{eq:data_vec_7.9}
    \textbf{d}_1 = \left(\begin{array}
    {c}\Delta^2_{21}(k_1, z_1)  \\ \Delta^2_{21}(k_2, z_1) \\ \vdots \end{array}\right),
     \textbf{d}_2 = \left(\begin{array}
    {c}\Delta^2_{21}(k_1', z_2)  \\ \Delta^2_{21}(k_2', z_2) \\ \vdots \end{array}\right),
\end{equation}
In general  $k_i \neq k_i^{\prime}$. Considering a single data vector, in order to reduce the 
covariance between the neighbouring $k$-bins, we follow the method adopted in Ref.~\cite{HERA:2021noe}. 
Starting from a minimum $k$-bin (which is decided on the basis of the signal-to-noise ratio), 
we include the alternate $k$-bins in our analysis such that the bins do not have significant overlap. 
In the following derivation, we show the form of the likelihood function for a single data vector,
and the composite likelihood for the two data vectors may be calculated as
$\mathcal{L} \propto \mathcal{L}_1 (\mathbf{d}_1) \cdot \mathcal{L}_2(\mathbf{d}_2)$.

Following Ref.~\cite{HERA:2021noe}, we assume a Gaussian likelihood for the data, 
and denote the unknown systematics as $\textbf{u}$, so that one can express 
the likelihood function for a data vector $\textbf{d}$ as 
\begin{align}
\label{eq:likelihood_func}
\mathcal{L}(\mathbf{d}|\theta,\mathcal{M},\mathbf{u})\propto \exp\left(-\frac{1}{2}r(\theta,\mathbf{u})^T\Gamma r(\theta,
\mathbf{u})\right)
\end{align}
where $r(\theta,\mathbf{u}) \equiv \textbf{d} - \textbf{W}\cdot\textbf{m}(\theta) -\textbf{u}$, $\textbf{m}(\theta)$
is the emulator prediction for the data vector $\textbf{d}$, $\textbf{W}$ is the Window function (which accounts for the point spreading in Fourier k space \cite{Tegmark:1996qt, Liu:2011hh, Dillon:2013rfa, Liu:2019awk, Cunnington:2019lvb}),
and $\Gamma \equiv \Sigma^{-1}$ where $\Sigma$ in the covariance matrix.
In the parameter estimation problem, we are interested in finding the posterior probability 
for obtaining a parameter set given the observation data. This 
can be expressed in terms of the likelihood, based on Bayes' theorem, as 
\begin{equation}\label{eq:Bayes}
    p(\theta| \textbf{d}, \mathcal{M}, \textbf{u}) \propto \mathcal{L}(\textbf{d} | \theta, \mathcal{M},\textbf{u}) p(\theta | \mathcal{M}) p(\textbf{u}),
\end{equation}
where $p(\theta | \mathcal{M})$ is the prior function of the parameter set $\theta$ given the model
$\mathcal{M}$, and $p(\textbf{u})$ denotes the priors on the systematics. 
In general, priors on systematics should come from the experiments. However, in case of HERA, the 
systematics are largely unknown. Therefore, we follow Ref.~\cite{HERA:2021noe} and 
marginalize over the systematics prior range in order to obtain the likelihood. We assume 
a multivariate uniform distribution for $\textbf{u}$, where the systematic noise in each $k$ and $z$
bins are independent. The marginalization can be done using,
\begin{align}
  \mathcal{L}(\mathbf{d}|\theta,\mathcal{M},\mathbf{u})\propto \int_{\textbf{u}_{\rm min}}^{\textbf{u}_{\rm max}} \exp\left(-\frac{1}{2}\textbf{r}(\theta,\textbf{u})^T \Gamma \textbf{r}(\theta,\textbf{u}) \right) d\textbf{u},\,.
    \label{eq:marg_like_integral}
\end{align}
Assuming $\Gamma$ to be diagonal, the integral takes a simple form
\begin{align}
  \mathcal{L}(\mathbf{d}|\theta,\mathcal{M},\mathbf{u})\propto  \prod_i^{N_{d}} \int_{u_{i, {\rm min}}}^{u_{i, {\rm max}}} \exp\left(-\frac{\left[y_i - u_i\right]^2}{2\sigma_i^2}\right) du_i,
    \label{eq:product_integral}
\end{align}
where the index `i' denotes the $k$ bin number, 
$y_i = d_i - [Wm(\theta)]_i$ is the difference between the model and the data at each bin,
$N_d$ is the number of k bins, and $\sigma _i$ are the diagonal elements of the covariance matrix. 
In order to calculate the integral, we set $\textbf{u}\in [0, \infty]$, assuming systematics to be 
positive (although negative systematics can occur in some cases \cite{Kolopanis:2019vbl} and 
those are subject to interpretation).
Marginalization over the systematics using the assumed bounds yields 
\begin{align}
\label{eq:final_marg_likelihood}
   \mathcal{L}(\mathbf{d}|\theta,\mathcal{M})\propto \prod_i^{N_{d}}\frac{1}{2}\left(1+{\rm erf}\left[\frac{y_i}{\sqrt{2}\sigma_i}\right]\right),
\end{align}
where `$\rm erf$' is the error function. 
We note that the marginalized likelihood function is sensitive to the upper bounds of the data points, which is
particularly useful for the HERA power spectrum data.

\subsubsection*{Additional Observables}\label{external_likelihoods}

Note again that HERA measurements place only an upper bound on the 21-cm 
power spectrum. Therefore, the Likelihood in Eq.~\eqref{eq:final_marg_likelihood}
does not impose any tight bounds on the model parameters. Following Ref.~\cite{HERA:2021noe}, we use 
additional observables from other experiments to complement the HERA power spectrum data. These 
additional observables are: (i) the upper bound on the neutral hydrogen fraction
$x_{H\RNum{1}}<0.06+0.05(1\sigma)$ at 
$z\sim5.9$, measured by the dark fraction on high redshift quasar spectra \cite{McGreer:2014qwa},
(ii) the faint galaxy
UV luminosity functions at $z=6, 7, 8, 10$ \cite{Bouwens:2014fua,Bouwens:2016aaa,Gillet:2019fjd} 
\footnote{https://github.com/21cmfast/21CMMC/tree/master/src/py21cmmc/data}, and
(iii) the Thomson scattering optical depth of CMB photons $\tau_e=0.0569^{+0.0081}_{-0.0066}$, using Planck~\cite{Planck:2018vyg} data from Ref.~\cite{Qin:2020xrg}. We shall see later that these additional observables help to place tight bounds on 
some of the model parameters, while HERA helps constrain some of the remaining parameters.

\section{Artificial Neural Network (ANN)}\label{section_3}

In order to perform a Markov chain Monte Carlo (MCMC) analysis, we need to be able to produce realisations of the 
21-cm power spectra and other global quantities, given a set of input parameters. 
However, to simulate the 21-cm signal for a large cosmological volume with the required resolution
using the latest version of \texttt{21cmFAST}~\cite{Munoz:2021psm}  takes $\mathcal{O}(1\,{\rm hour})$. 
The traditional MCMC pipeline requires to generate a large number of simulations in a sequential manner to be able to construct the final posterior, which takes a lot of time and can be prohibitive. In order to address this, we have developed an artificial neural network (ANN)-based
21-cm signal emulator, which we describe below.

\subsection{ANN Architecture}

\subsubsection{Simulations}

\begin{table*}[]
\begin{ruledtabular}
\begin{tabularx}{\textwidth}{ll}
{\bf Parameters} & {\bf Description} \\ \hline

$f_{\star}$ \dotfill &  Stellar to halo mass ratio for haloes \\ \hline

 $\alpha_{\star}$ \dotfill &  Stellar-to-halo mass power-law index \\ \hline

 $f_{\rm esc}$ \dotfill &  Escape fraction of ionizing photons from the haloes \\ \hline

 $\alpha_{{\rm esc}}$ \dotfill &  Escape fraction of ionizing photons to halo mass power-law indices \\ \hline

 $M_{\rm turn}$ \dotfill &  Mass scale below which inefficient cooling and/or feedback suppresses efficient star formation \\ \hline

 $t_{\star}$ \dotfill & Star formation time-scale in units of $H^{-1}(z)$ \\ \hline

 $\Lx$ \dotfill & Soft-band X-ray luminosity per SFR in units of ${\rm erg} \;{\rm s}^{-1}\;{\rm M}_{\odot}^{-1}\;{\rm yr}$
for ACGs \\ \hline

 $E_0$ \dotfill &  Minimum X-ray energy (in eV) escaping the galaxies into the IGM \\ \hline

 $\alpha_{{\rm X}}$ \dotfill &  Spectral index of X-ray sources \\

\end{tabularx}
\end{ruledtabular}
\caption{The main astrophysics parameters.}
\label{table:astro_params}
\end{table*}

It is relatively well accepted that the galaxies form inside the dark matter halos, with different 
feedback processes shape their evolution~
\cite{Silk:2013xca, Vogelsberger:2013eka, Somerville:2014ika, Pillepich:2017jle, Vogelsberger:2019ynw, Wetzel:2022man}. 
These galaxies are the main sources of various radiation fields that
affect the 21-cm signal from the beginning of cosmic dawn to the epochs that follow. We use the 
semi-numerical code \texttt{21cmFAST} that simulates various galaxy properties, depending on the host halo
mass, and produces these radiation fields
~\cite{Qin:2020xyh, Qin:2020pdx, Munoz:2021psm, Mesinger:2010ne, Murray:2020trn, Sarkar:2022mdz, Shmueli:2023box}. 
The code then uses the same radiation fields to deduce the 21-cm fluctuations at different redshifts. Note that the semi-numerical calculations of  
\texttt{21cmFAST}  are tuned to reproduce the galaxy UV luminosity functions
~\cite{Mason:2015cna, Park:2018ljd, Rudakovskyi:2021jyf} observed at reionization as well as
the observed opacity of IGM. 

Given a set of cosmological and astrophysical parameters,  \texttt{21cmFAST}  generates 
realizations of the 21-cm fluctuations in coeval and/or light-cone boxes. Those boxes are then used to
compute the global quantities like the mean brightness temperature of the 21-cm signal 
$\langle \delta T_{21} \rangle$, optical depth of reionization $\tau_{\rm reio}$, etc., and 
21-cm fluctuation power spectrum $\Delta^2_{21}(\mathbf{k})$. In addition to this, 
\texttt{21cmFAST} is also able to output the UV luminosity function of galaxies at various redshifts.
We create an ensemble of the observables (which we call data sets)
by changing the model parameters, and this ensemble will be used to
train and validate our emulator, as we describe later in this section. 
For the 21-cm power spectrum, in this work we are mainly concerned about two redshift bins $z=7.9$ and $10.4$
where HERA observations have the best detection~\cite{HERA:2021noe}. 

The definitions of the important astrophysical parameters used in this work are given in 
Table~\ref{table:astro_params}. The \texttt{21cmFAST} code models the effects
of both the PopIII and PopII stars. 
PopIII~\cite{Bromm:2003vv, Bromm:2005gs} refers to the first generation of stars believed to reside in 
molecular cooling halos (MCGs) or mini-halos, while
PopII~\cite{Smith:2015vpa} refers to the second generation of stars
which predominantly reside in atomically cooled galaxies or ACGs. 
We shall discuss these in more detail in Section~\ref{sec:role_of_mhs}. 
The parameters related to the ACGs (MCGs) are represented by a `10' (`7' or `mini') 
in the subscript (referring to the typical halo mass). In the next few sections, we shall assume
that the cosmic dawn and reionization processes are driven solely by the PopII stars.
We build the data set in the following section using \texttt{21cmFAST}  where we consider only
PopII stars. We note that the same modelling has been considered in Ref.~\cite{HERA:2021noe}.

\subsubsection{Building the Data set}
\label{sec:data_builing}

In this section, we discuss the method of composing the data set which will then be divided 
into training, validation and testing sets.
We start by determining the desired range over which we want to vary each parameter. Having its range fixed, 
we sample the whole parameter space 
using a Latin Hypercube (LH) sampler  which seeks
to produce uniform sampling densities when all points
are marginalized onto any one dimension \cite{McKay1979}. 
After getting the sampled parameter sets, we run the \texttt{21cmFAST} code to 
obtain the 21-cm power spectrum and global quantities like $\tau_e$, $x_{\rm HI}$  
for each combination. We then save these quantities along with the 
parameter values. 

We now discuss the data set which consists of the 21-cm power spectrum as a function of $k$. 
Note that, we have a limited number of $k$-bins from the simulations to begin with. 
Before the training process, we use interpolation to increase the number of  power spectrum $k$-bins to 70, 
ranging from $k = 0.1$ Mpc$^{-1}$ to $k = 1.73$ Mpc$^{-1}$, so that the variation with the wave number is smoother.
The data set is then classified into 6 different classes (denoted by index `n') 
according to the amplitude difference between k bins. 
The first 4 $(n = 0-3)$ classes are for signals where the difference between the maximal amplitude and the minimal amplitude of the power spectrum, is smaller than $10^n$ mK$^2$ but greater than $10^{(n-1)}$ mK$^2$. 
Class 5 is for signals where the difference between the maximum and minimum power spectrum values in 
different bins, i.e. [max(power spectrum) - min(power spectrum)] is $ > 10^3$ mK$^2$,
Finally, class 6 is reserved for  signals where the minimal amplitude is not in the lowest $k$ bin 
(these are exotic-shaped signals). This classification is done so that when the emulator is trained, 
we can take approximately the same number of samples from each class,
so that the trained emulator will not be biased towards any of the classes. Without doing so, emulating exotic signals tends to be more difficult, since they are rare in the un-manipulated data set. 
Note also that the `class 0' signals are also very hard to emulate even after this process because 
most of them are actually `dead' or flat signals having no features that the emulator can learn. 
This happens when for some parameter combination, the fluctuations at all scales cease at a specific 
redshift. For example, when the IGM becomes sufficiently ionized at low redshifts that the fluctuations 
in the electron density field become more or less uniform, and the power spectrum is flat (and even becomes zero 
in some cases). Emulating the flat signals is extremely difficult, and we have 
decided not to include these at all in the emulating process. Including flat signals in the emulating process 
leads to unreliable predictions. A method for predicting these flat
signals (for use in the MCMC pipeline) will be explained later. The bottom line is that we have discarded the class 0 signals when  training the 
emulator.

For the training process, we take $\sim$ 2500 randomly sampled signals from each class for $z =7.9$, and $\sim$ 3000 signals from each class for the $z=10.4$ band. We choose the numbers such that we have a similar number of samples
in each class, and such that no class is over/under-represented. 
Finally, the data set is divided into training (85\%), validation (10\%) and testing (5\%) sets.
Before the training process, the astrophysical parameters are normalised to be in the range $[-1,1]$.
We do not normalize the power spectrum as our initial tests suggested a better performance with the 
original power spectrum.

\subsubsection{NN Architecture}

In this section, we discuss the architecture of our NN emulators. 
All of our NN emulators and classifiers are built using the \texttt{TensorFlow} \cite{tensorflow2015-whitepaper} and \texttt{Keras} \cite{chollet2015keras}
libraries defined in \textsc{Python3}.
Here we consider a 9-parameter emulator model based on \texttt{21cmFAST}.
According to this model, 
only the halos which are cooled by atomic line transitions
host sources that start cosmic dawn and cosmic reionization. 
The choice of the parameters is based on Ref.~\cite{HERA:2021noe}. As explained above, 
there is another possible cooling mechanism, 
which makes it possible to form sources inside mini-halos. 
Using \texttt{21cmFAST} simulations, it is possible to model these mini halos, and we 
shall discuss this in more detail in the later sections. 

Therefore, the emulator has 9 input parameters, 70 output $k$ bins, and 6 fully connected hidden layers 
- of 288, 512, 512, 288, 512, 1024 dimensions, respectively. The number of layers and the number of neurons 
in each was chosen after an optimization process. We have tested the emulator for a number of 
conventional activation functions, namely ReLU \cite{4082265} and $tanh$, etc. We finally settled with the 
following function (taken from  Ref.~\cite{SpurioMancini:2021ppk}) 
\begin{equation}\label{activation_func}
    f( \textbf{x}, \bm{\alpha}) = \textbf{x} \odot \left ( \bm{\alpha} + (1-\bm{\alpha}) \odot \left ( 1+e^{-\textbf{x}} \right )^{-1} \right )\,,
\end{equation}
where $\textbf{x}$ are the layer inputs, $\bm{\alpha}$ is a vector of trainable parameters, 
and $\odot$ denotes the element-wise product. The activation is followed by a batch normalization layer~\cite{Ioffe:2015ovl}, 
which standardises the input for the next layer. After the forward propagation, the loss is calculated using a relative percentage error  
\begin{equation}\label{loss_func}
    \mathcal{L}(y_{true}, y_{pred}(\theta)) = \frac{|y_{true} - y_{pred}(\theta)|}{y_{true}}\times 100\,.
\end{equation}

\begin{figure}[h]
    \centering
\includegraphics[width = \columnwidth]{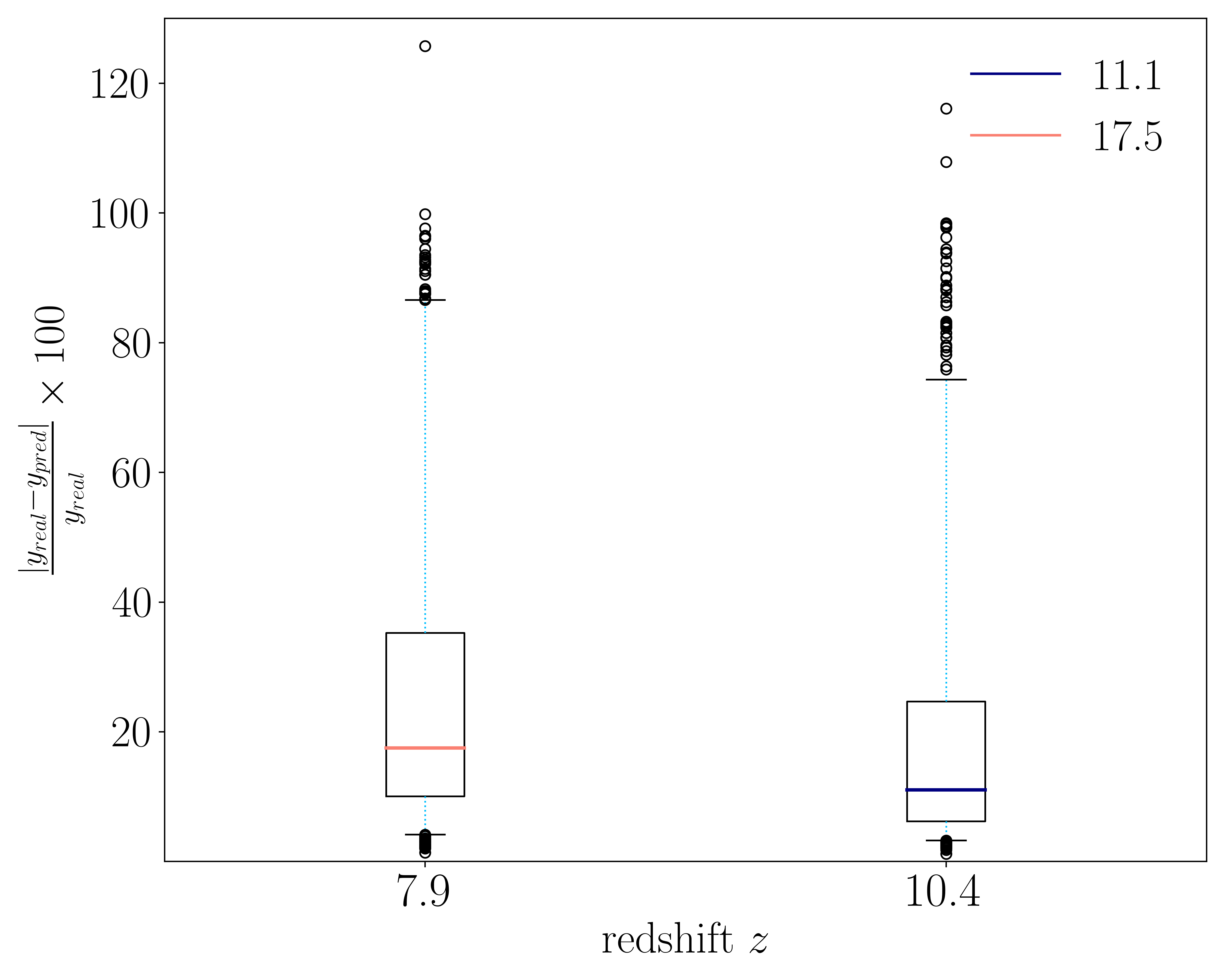}
    \caption{Emulator loss statistics on the testing sets, of size $\sim$ 650 signals for each band. The box represents all the samples that fall between the 25th and the 75th percentile, and the line in the middle of it is the median. The black circles are outliers that extend beyond the 95th percentile.}
    \label{fig:emulation_results}
\end{figure}

The training process is performed using mini batches of size 128 samples, so that the trainable parameters are updated using the gradients of the loss function, after forward propagating each batch. The training process was done using the Adam \cite{Kingma:2014vow} optimizer, where the initial learning rate is 0.01, and is reduced when the prediction on the validation set does not improve over more than 10 epochs. The training phase ends when there is no improvement over more than 30 epochs, and the best weights are stored at that moment. Finally, we examine our emulator performance on the test set, and the results are presented in Fig.~\ref{fig:emulation_results}.

\subsubsection{Retraining}

After training the emulator, we review its performance very carefully on the test set. 
In the context of the power spectrum, we find that on rare occasions (less than 5\%), 
the emulator prediction differs from the real value by almost 100\%. This may bias 
the MCMC results, particularly the median value. 
We do the following to make sure that this error does not affect our results. 

After the first training, we determine a rough $\sim 1\;\sigma$ region in the parameter space 
by running the MCMC analysis using the emulator. We then generate $\sim 5000$ more signals 
by sampling the rough $1\;\sigma$ parameter space using the LH method, where the $1\;\sigma$ is considered for all the parameters except $\Lx$, for which the emulator has to be accurate  over the full prior range. We then re-train our emulator 
on this re-sampled data set. 

\begin{figure}[h]
    \centering
    \includegraphics[width = \columnwidth]{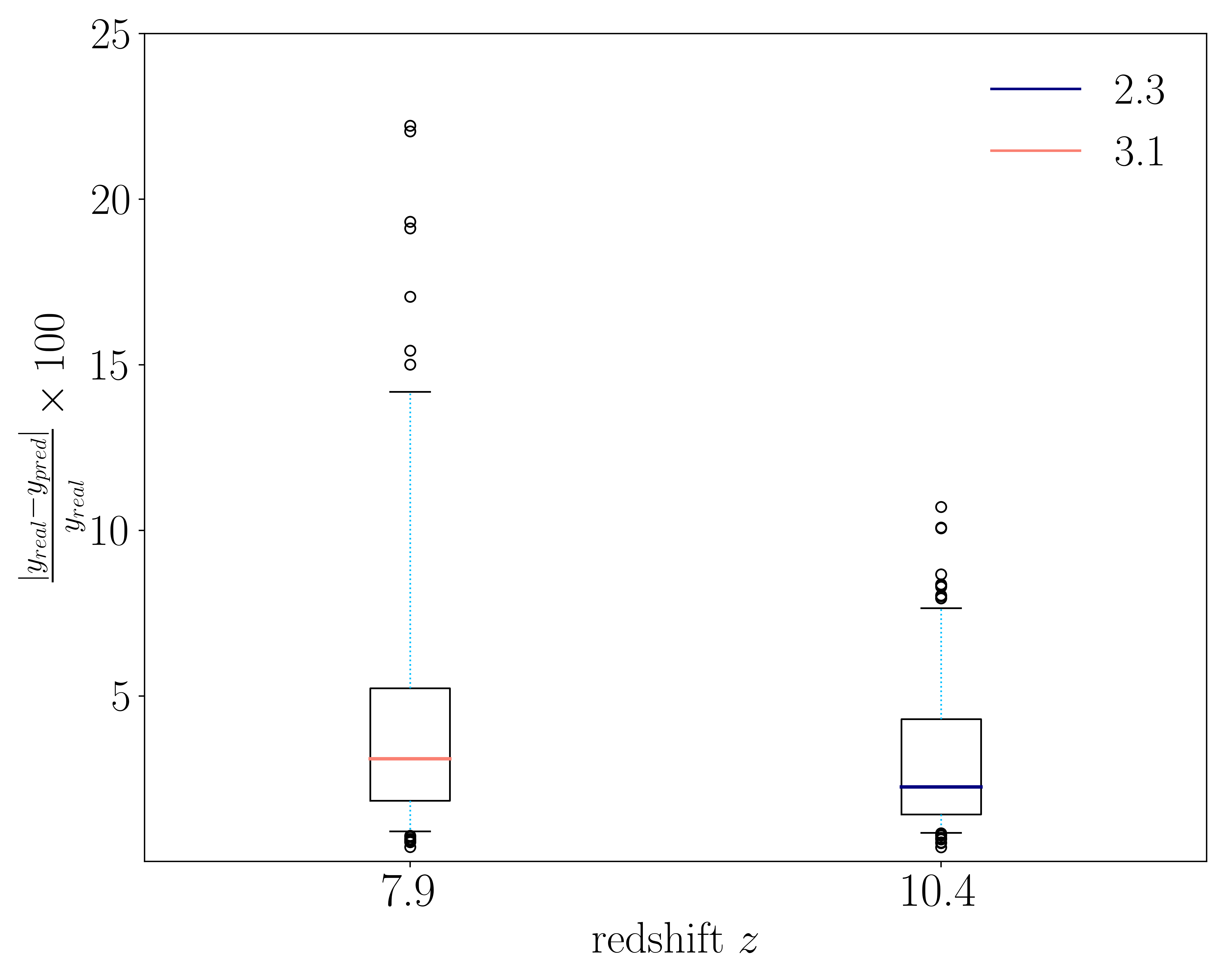}
    \caption{Retrained emulator loss statistics on our testing sets, of size $\sim$ 250 signals for each band. The box represents all the samples that are between the 25th and the 75th percentiles, and the line in the middle of it is the median. The black circles are outliers that extend beyond the 95th percentile.}
    \label{fig:my_label}
\end{figure}

\begin{figure}[h]
    \centering
\includegraphics[width = \columnwidth]{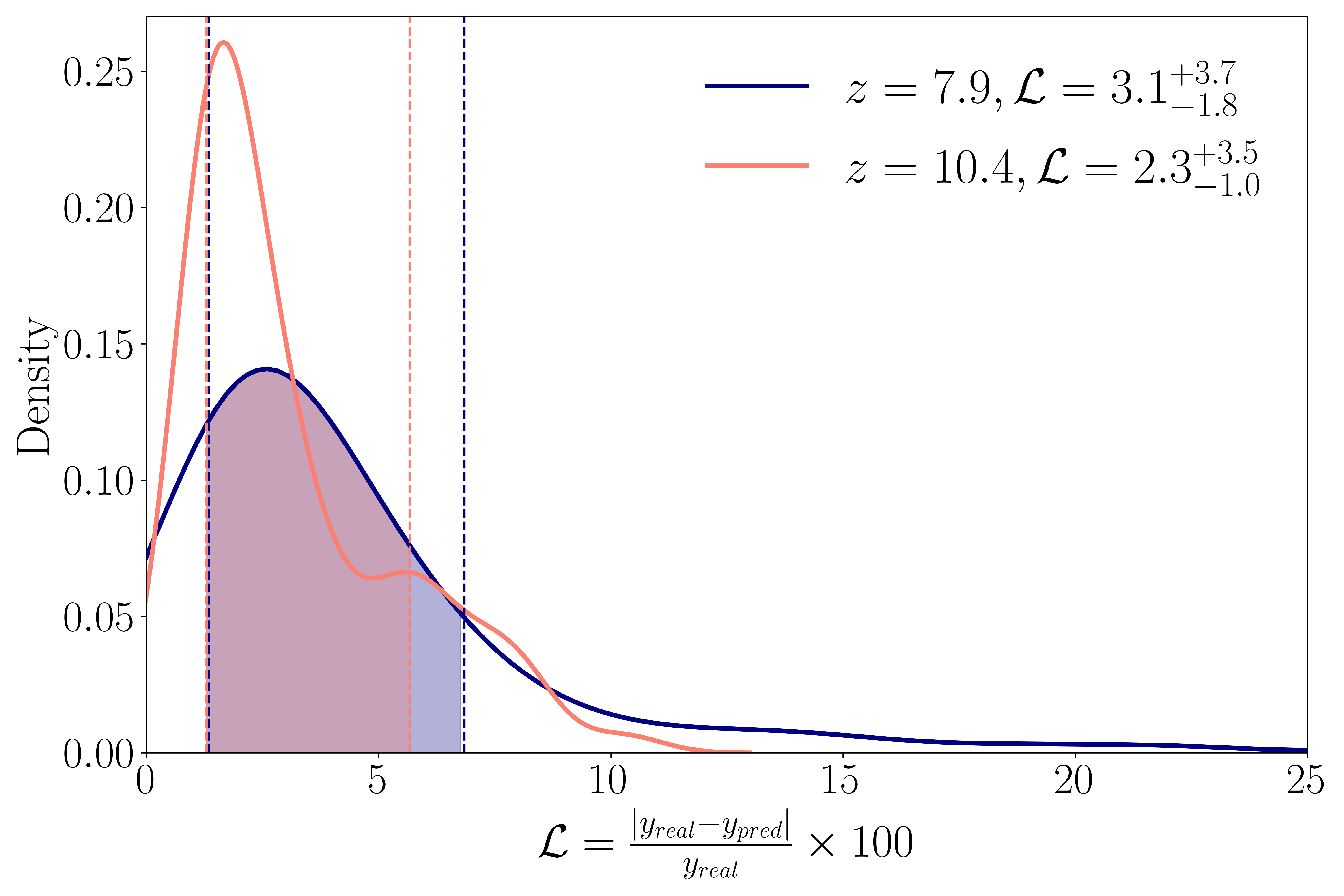}
    \caption{Different view of the loss statistics for the retrained emulator. The dashed lines represent the 16th and 84th percentiles respectively, and their exact value is written along with the median value in the figure legend.}
    \label{fig:retraining_results}
\end{figure}

Note that, for some astrophysical parameters such as $\alpha_{\ast}$,
the $1\;\sigma$ constraint range is very small  as we shall show later in Fig.~\ref{fig:mcmc_results}. For those parameters, we choose a wider range when sampling. 
We show the performance results after the re-training process in Figs.~\ref{fig:my_label} and \ref{fig:retraining_results}. 
We see that the re-trained emulator gives a median error of 2-3\% in the $1\;\sigma$ parameter space.
Now considering the whole range of parameters that we used for initial training, we see that our 
re-trained emulator exhibits $<20\%$ median error outside of the $1\;\sigma$ parameter space. 
This suggests that the emulator will have better precision, at least 
within the $1\;\sigma$ parameter space, such that it will not bias the MCMC analysis by much and will
provide better $1\;\sigma$ constraints. 

 We shall see later (see also Ref.~\cite{HERA:2021noe})
that most of the astrophysical parameters, 
courtesy of the additional observables used in addition to the HERA power spectrum data, are very well constrained.
Other parameters, like $\Lx$ (which is meaningfully constrained only at the lower side) or 
$E_0$, have almost no constraints from the observations. Now, due to the property of the Markov chains, 
the MCMC algorithm will mostly sample the region of the parameter space where our retrained emulator performs extremely well. 
This means we shall have many accurate MCMC samples in the vicinity of the $1\;\sigma$ range,
and very few less-accurate samples outside this region. This weighs down any possible bias and ensures our predictions
are robust.

\subsubsection{Classification}
\label{sec:classification}

We have already mentioned in Section~\ref{sec:data_builing} 
that the class 0 signals or flat power spectra are very hard to emulate. 
The power spectrum amplitude in this class is close to zero, $\lesssim 3$ mK$^2$ for all the  k bins.
When we run the MCMC process, for each set of parameters we need to know whether 
the parameters will produce flat signals even before we feed those to the ANN.
We, therefore, need machinery that predicts the occurrence of the flat signal
for a given parameter set. 

We have built a neural network classifier for the above purpose. 
This classifier is trained at each frequency band to predict the occurrence of the 
flat 21-cm power spectrum. We discuss the architecture of the classifier in Appendix~\ref{apndx_c}. 
We find that the classifier's prediction accuracy is $\gtrsim 97\%$. This indicates that, on average, 
we shall have a miss-classification less than once in every 30 MCMC iterations. This does not 
boast a serious error in the MCMC analysis.

Considering the flat signal, we choose the following path. When the classifier predicts a flat signal for
some parameter combination, we generate a random power spectrum at each $k$-bin where 
the amplitude is drawn from a Gaussian distribution having a mean of 2 mK$^2$ and a standard deviation
of 1 mK$^2$. This we do irrespective of the parameter values. Our goal here is not to predict the flat signal 
accurately as these signals are still somewhat rare. Also, considering the upper bound on the 21-cm power spectrum 
from HERA (and our chosen likelihood function in Eq.~\eqref{eq:final_marg_likelihood}) 
it is clear that any flat signal below the upper limit is allowed. Therefore, HERA data does not really discard these 
parameter values, rather the constraints come from the additional likelihoods. Therefore, we surmise that 
our solution for tackling the flat power spectra does not pose any serious issues.

\subsection{Emulating the other observables}

We extend our NN for emulating $x_{HI}$ and $\tau_e$. These are easy to emulate as 
for a fixed redshift both of them are just single numbers. We have therefore used a relatively simple neural network architecture  for emulating both $x_{H_I}$ and $\tau_e$. As we discuss in Appendix \ref{apndx_1}, we find that the emulators performs extremely well and the 
emulation error is negligible. As for the simulated UV luminosity functions, we do not have 
any emulators for these. UV luminosity functions can be calculated analytically very quickly via \texttt{21cmFAST}, and in the MCMC analysis we generate these directly by plugging in the parameters.

\section{The role of Mini-Halos}\label{sec:role_of_mhs}

The current understanding of the stellar composition in galaxies at low redshifts is quite comprehensive.
However, our knowledge about the initial galaxies that sparked cosmic reionization is limited. According to the hierarchical model of structure formation, the first generation of stars, known as 
PopIII stars, 
likely formed inside small molecular cooling galaxies or MCGs around $z\sim20-30$~
\cite{Haiman:1995jy, Haiman:1996rc, Tegmark:1996yt, Abel:2001pr, Bromm:2003vv, 
Yoshida:2006bz, Haiman:2006si, Trenti:2010hs}. 
Inside MCGs, the gas cools mainly
through the ${\rm H_2}$ rotational–vibrational transitions efficient at
$T_{\rm vir} \sim 10^3 - 10^4\,{\rm K}$.
These early galaxies were believed to be hosted by halos with virial temperatures $\lesssim 10^4$ K, 
corresponding to total halo masses of $\lesssim 10^8\,{\rm M}_{\odot}$ during the epochs of reionization and cosmic dawn
~\cite{OShea:2007ita,Mirocha:2017xxz,Trenti:2010hs,Qin:2020xyh,Munoz:2021psm}.
These small halos are often called \textbf{mini-halos}. Over time, feedback mechanisms suppressed 
star formation in these MCGs, leading to the emergence of heavier, atomically cooled galaxies (ACGs) 
with halo masses $\gtrsim10^8\,{\rm M}_{\odot}$. Therefore, 
most of the ACGs at high redshifts are ``second-generation'' galaxies, forming out of MCGs.
It is believed that the second generation of stars, known as PopII stars, predominantly
reside in ACGs~\cite{Qin:2020pdx}.

PopII stars likely drove most of the heating and ionization process 
at $z\sim5-10$~\cite{Mason:2019ixe,Choudhury:2020vzu,Visbal:2015rpa}. 
However, there should be a small but significant contribution from the PopIII
stars or MCGs as well. The star formation in MCGs is subject to  a number of feedback processes like
the Lyman-Werner radiation feedback~\cite{Munoz:2019rhi, Fialkov:2012su, Visbal:2014fta, Safranek-Shrader:2012zig, Ricotti:2000at, Haiman:1996rc, Ahn:2008uwe}, 
baryon supersonic streaming velocities~\cite{Fialkov:2014rba, Barkana:2016nyr, Tseliakhovich:2010bj, Bovy:2012af, Stacy:2010gg, Fialkov:2011iw, Schmidt:2016coo},
etc. Depending on these, the radiation contribution
from the MCGs varies. MCGs spark the cosmic dawn epoch earlier compared to the scenarios where only ACGs exist
~\cite{Munoz:2019rhi}. 
X-ray photons from MCGs also prepone the epoch of heating~\cite{Munoz:2019rhi}, although MCGs are believed to have a 
negligible contribution to reionization. As emphasized above, in Ref.~\cite{HERA:2021noe}, the 21-cm signal was modelled 
without taking the contribution of the PopIII stars or MCGs into account. One of the main conclusions
of Ref.~\cite{HERA:2021noe} is that the early galaxies were more X-ray luminous than their present-day
counterparts. Our main goal in this paper is to check the validity of this result in the presence of 
a contribution to the 21-cm signal from 
PopIII stars.

One of the most salient features in the latest version of \texttt{21cmFAST} code~\cite{Munoz:2021psm} is that it 
now includes the contributions of both the MCGs and ACGs. The code assumes MCGs host 
PopIII stars with a simple stellar-to-halo mass relation (SHMR), distinct from that of ACGs hosting 
PopII stars. For a comprehensive reading about how \texttt{21cmFAST} incorporates 
PopII and PopIII stars in the calculations, the reader is referred to Ref.~\cite{Munoz:2021psm, Qin:2020xyh}.

As mentioned earlier, star formation in mini haloes depends on feedback processes which
further decide the molecular-cooling turnover halo mass. Haloes below this mass
are not capable of forming stars. In \texttt{21cmFAST}, this turnover mass is a product of three
factors \cite{Munoz:2021psm},
\begin{equation}
    M_{\rm mol} = M_0 (z) f_{v_{\rm cb}} (v_{\rm cb})f_{\rm LW}(J_{21}) \,,
\end{equation}
where $M_0 (z)$ is the molecular cooling threshold in the absence of feedback with
$M_0 (z)=\Tilde{M_0}\times(1+z)^{-3/2}$ where $\Tilde{M_0}=3.3\times10^7\,{\rm M}_{\odot}$
(corresponding to a viral temperature of $10^3$ K \cite{Tegmark:1996yt}), 
$f_{v_{\rm cb}} (v_{\rm cb})$
and $f_{\rm LW}(J_{21})$ encodes the effects of respectively the baryon streaming velocity and
Lyman-Werner background radiation. 
$f_{\rm LW}(J_{21})$ is parametrized as,
\begin{equation}
    f_{\rm LW}(J_{21}) = 1 + A_{\rm LW} (J_{21})^{\beta_{\rm LW}}\,,
\end{equation}
with $A_{\rm LW}$ and $\beta_{\rm LW}$ are two free parameters that quantifies the strength
of the feedback, and $J_{21}$ being the Lyman-Werner radiation intensity. 
Similarly, $f_{v_{\rm cb}} (v_{\rm cb})$ is parametrized as,
\begin{equation}
    f_{v_{\rm cb}} (v_{\rm cb}) = 
    \left( 1 + A_{v_{\rm cb}} \frac{v_{\rm cb}}{v_{\rm rms}} \right)^{\beta_{v_{\rm cb}}}\,,
\end{equation}
where $v_{\rm cb}$ is the local value of the streaming velocity, 
$v_{\rm rms}$ is the rms of the streaming velocity, $A_{v_{\rm cb}}$ and $\beta_{v_{\rm cb}}$
are two free parameters that determine the strength of the feedback. We have kept the 
feedback parameter values fixed to 
$A_{\rm LW}=2.0$, $\beta_{\rm LW}=0.6$, $A_{v_{\rm cb}}=1$ and $\beta_{v_{\rm cb}}=1.8$
throughout all of the simulations that include MCGs.
Note that the feedback, or in other words, the parameters, are largely unknown at the
redshifts that we are interested in. The contribution of MCGs in cosmic dawn 
and reionization vary with the variation in these parameters. We do not vary these
in our analysis as these would increase the number of parameters. However, we expect that
our results will still be valid even when these parameters are varied in the vicinity of 
the fiducial values.

The \texttt{21cmFAST} code has a number of astrophysical parameters to characterize the galaxy properties. 
In Table~\ref{table:astro_params}, we mention some of the important astrophysical parameters used in this work.
We follow the same method to build the data set and utilize the same pipeline developed 
in Section~\ref{section_3} to emulate the 
21-cm power spectrum and other observables in the presence of MCGs. 
We discuss the performance of this emulator in Appendix \ref{apndx_b}. Overall, we find 
that we get similar results even after introducing the additional parameters for MCGs.

\begin{figure}[h]
    \centering
    \includegraphics[width = \columnwidth]{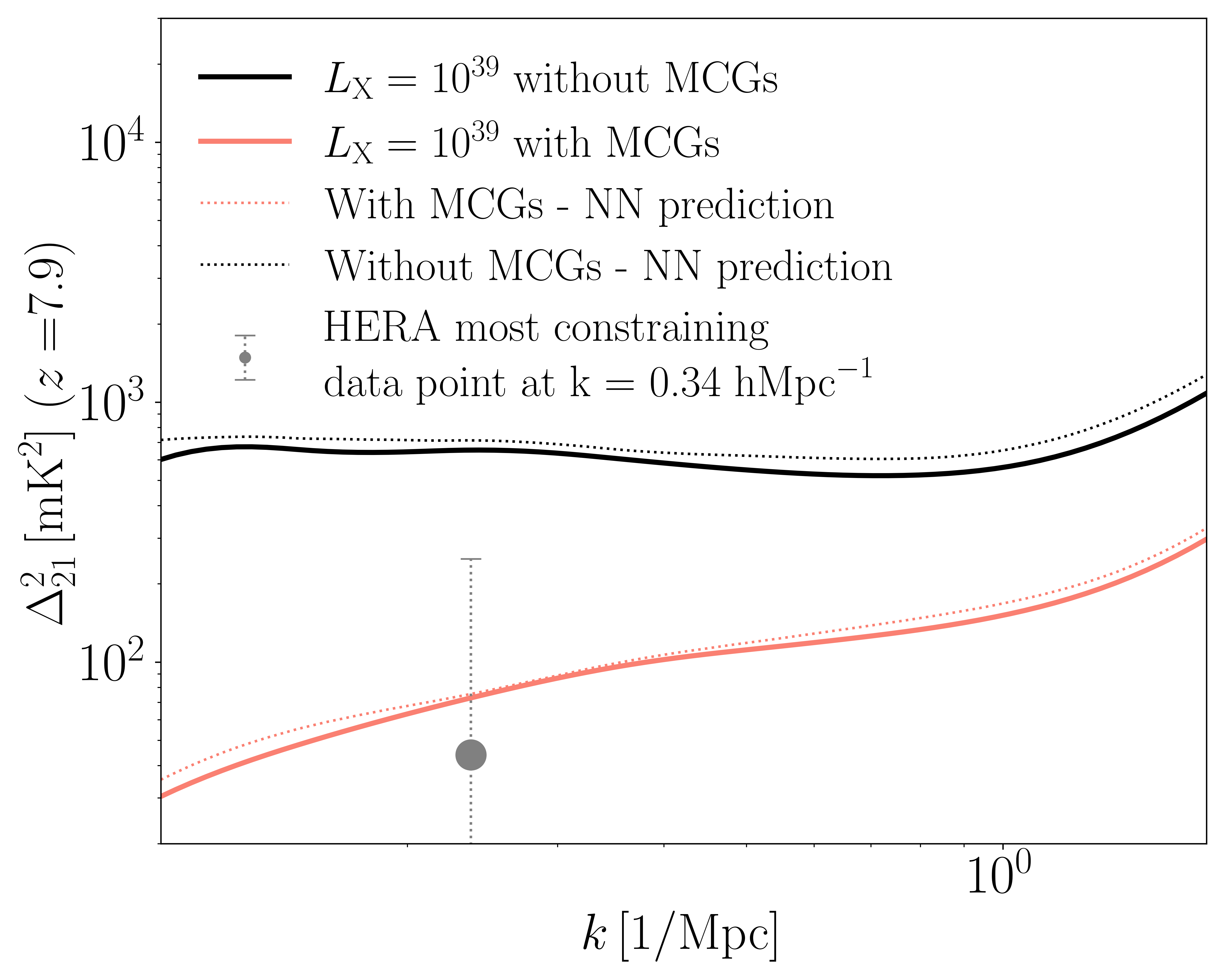}
    \caption{Comparison between simulated power spectra with and without MCGs, and the emulator performance for each signal. The parameters taken for the simulation are the medians of the marginalized posterior distributions depicted in Fig.~\ref{fig:mcmc_results_mini_halos}, for all the parameters (MCG parameters were not included in the ``without MCGs" simulation) except $L_{\rm X<2 \, keV}/{\rm SFR}$, for which we have taken $10^{39} \rm erg\, s^{-1}\,M_{\odot}^{-1}\,yr$.
    We quote some of the other parameters used in this figure for convenience: 
    $\log_{10} f_{\ast,10}=-1.25$, $\log_{10} f_{\ast}=-2.65$, $\alpha_{\ast}=0.52$,
    $\alpha_{\ast, {\rm mini}}=-0.03$, $\log_{10} f_{\rm esc,10}=-1.45$, 
    $\log_{10} f_{\rm esc,7}=-2.31$, $\alpha_{\rm esc}=0.55$. Note that, for 
    the ``without MCGs" scenario (red curves), we have considered only ACGs in the 
    simulations (same as the scenario assumed in Fig.~\ref{fig:mcmc_results}) with the 
    appropriate parameter values as mentioned above, 
    ignoring any contribution from the MCGs.}
    \label{fig:Lx=38_mini}
\end{figure}

In Fig.~\ref{fig:Lx=38_mini} we show the importance of accounting for the MCGs when interpreting the HERA
data. plotting the power spectrum for $\Lx = 10^{39} \Lu$ where we use only ACGs in the 
simulation, we see that this value of $\Lx$ is clearly disfavoured by the HERA 
observations. This is consistent with the findings in Ref.~\cite{HERA:2021noe}.
However, when adding MCGs to our simulations, we find that the same value of $\Lx$ is  consistent with the HERA data. A prediction for this behaviour, was already mentioned in Ref.~\cite{HERA:2022wmy}.
The bottom line is that with MCGs, even the smallest values of 
$\Lx$ in our prior range cannot be ruled out by current HERA data. Later, we shall see
the same in the posterior distribution of $\Lx$ from our full MCMC analysis. 

\section{Results}\label{section_5}

\subsection{Bayesian inference}

We now perform the MCMC analysis based on our pipeline that combines the emulator discussed in
Section~\ref{section_3} and the likelihood mentioned in Section~\ref{section_2}. In the MCMC analysis, 
we have two main goals:
(I) To reproduce the results of Ref.~\cite{HERA:2021noe} which includes only ACGs in the analysis, and
(II) To study the impact of MCGs on the X-ray heating constraints. 
Considering the two goals, we have two scenarios for the likelihood:
(A) Use a likelihood for only the additional observables mentioned in Section~\ref{external_likelihoods}, which
we denote as `\textbf{External likelihood}', and 
(B) Consider a `\textbf{Full likelihood}' that consists of the external likelihood and the likelihood for the
HERA power spectrum upper bound. We, therefore, have four different MCMC inferences and
these are:
(I-A): \textbf{ACGs + External likelihood},
(I-B): \textbf{ACGs + Full likelihood},
(II-A): \textbf{MCGs + External likelihood}, and
(II-B): \textbf{MCGs + Full likelihood}.
We discuss the results below. For the MCMC runs, we use the \texttt{Emcee} \cite{Foreman-Mackey:2012any} sampler. 
In table \ref{tab:prior_table}, we show the prior ranges used for this work and we use
uniform flat priors for all the variables. 

For inferences of type I-A and I-B, we have used the same
prior range  as in Ref.~\cite{HERA:2021noe}. 
For inferences with mini halos, we drop some parameters 
in order to reduce the parameter space. We find that $t_{\ast}$  (0.5) and
$\alpha_{X} $ (1.0) parameters are not very well constrained. Therefore, even if we drop them, 
there should not be any major effect on the inference. We keep them fixed at the 
values mentioned within the brackets. 
Furthermore, in principle,
Pop\RNum{3} stars should have different X-ray luminosity (denoted by $L^{\rm (III)}_{\rm X<2 \, keV}/{\rm SFR}$)
than Pop\RNum{2} stars (denoted by $L^{\rm (II)}_{\rm X<2 \, keV}/{\rm SFR}$). However, we have no prior
knowledge of $L^{\rm (III)}_{\rm X<2 \, keV}/{\rm SFR}$ of the Pop\RNum{3} stars, which reside mainly in MCGs. 
Considering this uncertainty, and, in order to reduce the number of parameters, 
we assume $L^{\rm (III)}_{\rm X<2 \, keV}/{\rm SFR}$ = $L^{\rm (II)}_{\rm X<2 \, keV}/{\rm SFR}$ 
and call this the {\it effective} $\Lx$. 
The prior range of the MCG parameters is taken to be the same as the ACG parameters.

Note that the above choice may affect the prior volume non-trivially.
Making $L^{\rm (III)}_{\rm X<2 \, keV}/{\rm SFR}$ = 
$L^{\rm (II)}_{\rm X<2 \, keV}/{\rm SFR}$ automatically reduces the effective value of 
$\Lx$. This would not be the case if $L^{\rm (III)}_{\rm X<2 \, keV}/{\rm SFR}$ is made
to vary independently. However, as the Pop\RNum{3} stars are metal-poor, 
our naive expectation is that 
$L^{\rm (III)}_{\rm X<2 \, keV}/{\rm SFR} \gtrsim L^{\rm (II)}_{\rm X<2 \, keV}/{\rm SFR}$.
Therefore, we expect our results to be consistent and even somewhat conservative under the assumption we made.

\begin{table}[h!]
\centering
    
\begin{tabular}{ |c|c|c|}
 \hline
 \multicolumn{3}{|c|}{Priors} \\
 \hline
 Parameter name &Lower bound &Upper bound\\
 \hline
 $\log_{10}f_{\ast,10}$   & -3.0     &  0.0\\
 $\log_{10}f_{\ast,7}$&   -3.5  & -1.0 \\
 $\alpha_{\ast}$ &-0.5 & 1.0\\
 $\alpha_{\ast,\rm mini}$    &-0.5 & 0.5\\
 $\log_{10}f_{{\rm esc},10}$&   -3.0  & 0.0\\
 $\log_{10}f_{{\rm esc},7}$& -3.0  & 0.0  \\
 $\alpha_{\rm esc}$& -1.0  & 1.0\\
 $\log_{10}[M_{\rm turn}/M_{\odot}]$ & 8.0 & 10.0\\
 $t_{\ast}$ & 0.0 & 1.0\\ 
 $\log_{10}\frac{L_{\rm X<2 \, keV}/{\rm SFR}}{\rm erg\, s^{-1}\,M_{\odot}^{-1}\,yr}$ & 38.0  & 42.0\\
 $E_0/{\rm keV}$ & 0.1 & 1.5 \\
 $\alpha_{X}$ & -1.0 & 3.0\\
 \hline
 
\end{tabular}

 \caption{Prior range summary for all the astrophysical parameters used in the different MCMC runs. Note that only a subset of them is used in each run, as explained in the text.}
\label{tab:prior_table}
\vspace{-0.15in}
\end{table}

\subsection{Weaker constraints on $L_{\rm X}$}

\begin{figure}[h]
    \centering
\includegraphics[width = \columnwidth]{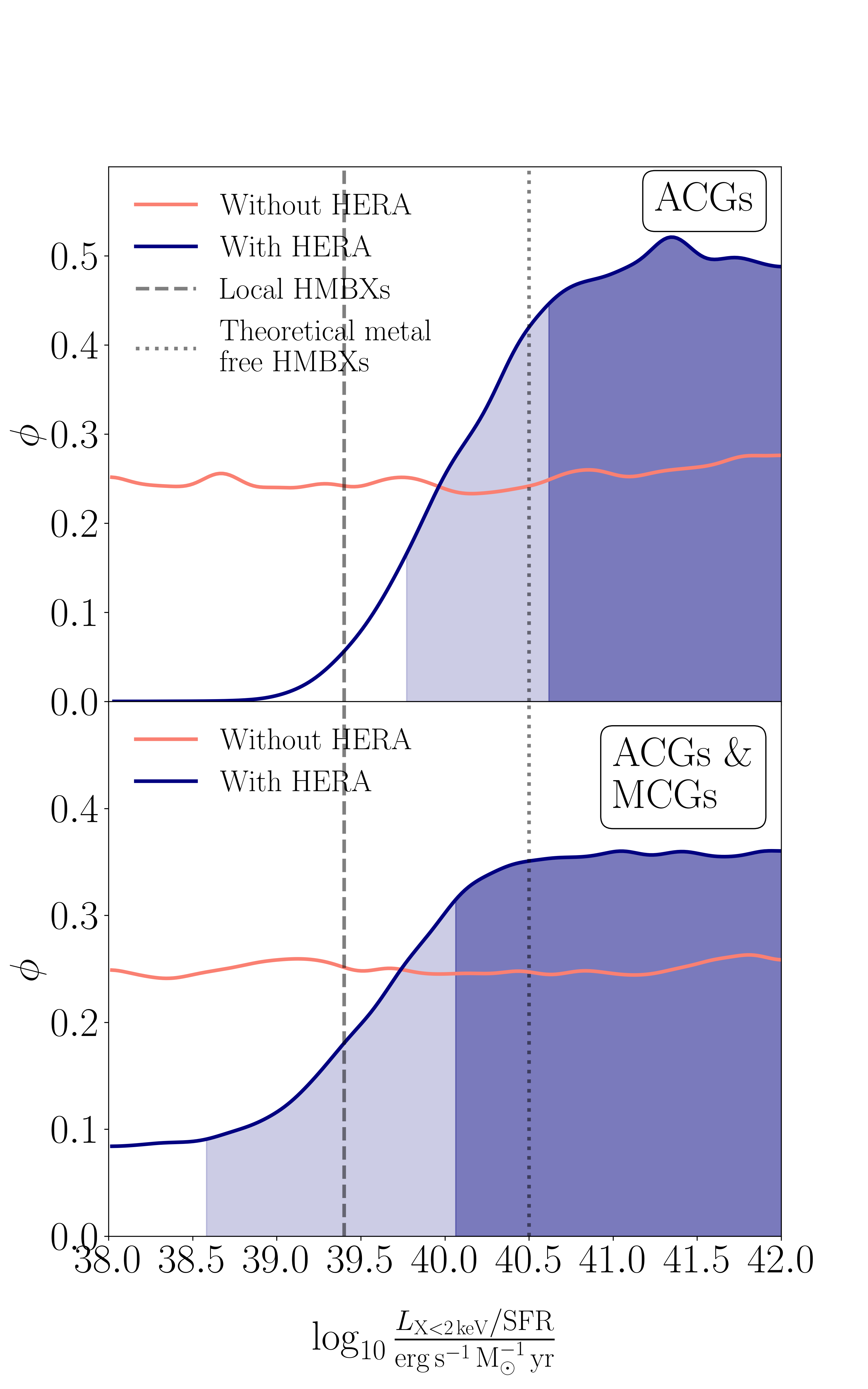}
\vspace{-0.15in}
    \caption{$L_{\rm X<2 \, keV}/{\rm SFR}$ marginalized posterior distribution. The dark shaded area denotes 68\% highest posterior density (HPD), and the light shaded area denotes 95\% HPD. The dashed black line represents the average $L_{\rm X<2 \, keV}/{\rm SFR}$ as inferred from  observations of high-mass X-ray binaries (HMXBs) in local, star-forming galaxies~\cite{Mineo:2011id}. The dotted black line denotes the theoretical prediction for metal-free HMXBs presented in Ref.~\cite{Fragos:2012vf} (such galaxies are expected to be common in the early Universe). The average observed value is outside the 2$\sigma$ region of the posterior distribution in the top plot, suggesting a possible tension between the observations. However, the bottom plot implies that this tension is relaxed when including MCGs in the 21-cm simulations.} 
\label{fig:Lx_marginalized}
\end{figure}

In the top panel of Fig.~\ref{fig:Lx_marginalized}, we show the marginalized 
posteriors for $\Lx$, the parameter most constrained by HERA,
for inferences I-A and I-B. This figure is directly comparable to 
Fig.~28 of Ref.~\cite{HERA:2022wmy} (see also Fig.~7 in Ref.~\cite{HERA:2021noe}) and our result is largely in agreement with theirs. 
The external likelihoods do not impose any constraint on $\Lx$. The HERA data
helps to place bounds at low $\Lx$, and we can see that values below 
$\Lx \sim 10^{40} \Lu$ are heavily disfavoured, while higher values are preferred. 

The amplitude of the power spectrum
increases as $\Lx$ is decreased, and below $L_{\rm X<2 \, keV}/{\rm SFR} \sim 10^{39} \Lu$
(as can be seen from Fig.~\ref{fig:Lx=38_mini}, where we change the $\Lx$ values
while keeping the other parameters at the median values as mentioned in Fig.~\ref{fig:mcmc_results}),
the power spectrum amplitude exceeds the HERA $1\;\sigma$ upper bound. On the other hand, we find 
an almost flat posterior towards the large $\Lx$. This result somewhat disagrees 
with Ref.~\cite{HERA:2022wmy} where the $\Lx$ posterior shows a steady decline
at $\Lx > 10^{40.8} \Lu$. This is possibly due to the unequally-weighted sample points of the MultiNest sampler ~\cite{Feroz:2008xx}
used by \texttt{21cmMC} ~\cite{Conversation}. The key point in 
Fig.~\ref{fig:Lx_marginalized} is that for models where only ACGs (or PopII stars) are 
present, the HERA upper limit suggests more X-ray luminosity for the high-redshift ($z>6$) sources.

The bottom panel of Fig.~\ref{fig:Lx_marginalized} shows the marginalized 
posteriors for $\Lx$, for inferences II-A and II-B. 
As discussed earlier, both these inferences are based on models where we include 
MCGs (or PopIII stars) and ACGs (or PopII stars), and $\Lx$ here represents an effective
value for the X-ray luminosity of these galaxies. The External likelihoods do not place any bounds on 
$\Lx$ within the prior range we used, which is expected. 
However, with HERA data, we find a striking difference when compared with the top panel. 
We see that the strong preference for the high $\Lx$ 
($\Lx > 10^{40} \Lu$) values is severely relaxed as soon as we include MCGs. 
This is also evident from Fig.~\ref{fig:Lx=38_mini}.
Although there is a decline in posterior probability below 
$\Lx \sim 10^{39.5} \Lu$, this is not as significant as in the top panel 
where the decline is very sharp at $\Lx < 10^{40} \Lu$. 
Overall, Fig.~\ref{fig:Lx_marginalized} clearly indicates that the lower $\Lx$ 
values are still possible when we add MCGs to the analysis.

\subsection{Posteriors for other astrophysical parameters}

\begin{figure*}
    \centering
\includegraphics[width = \textwidth]{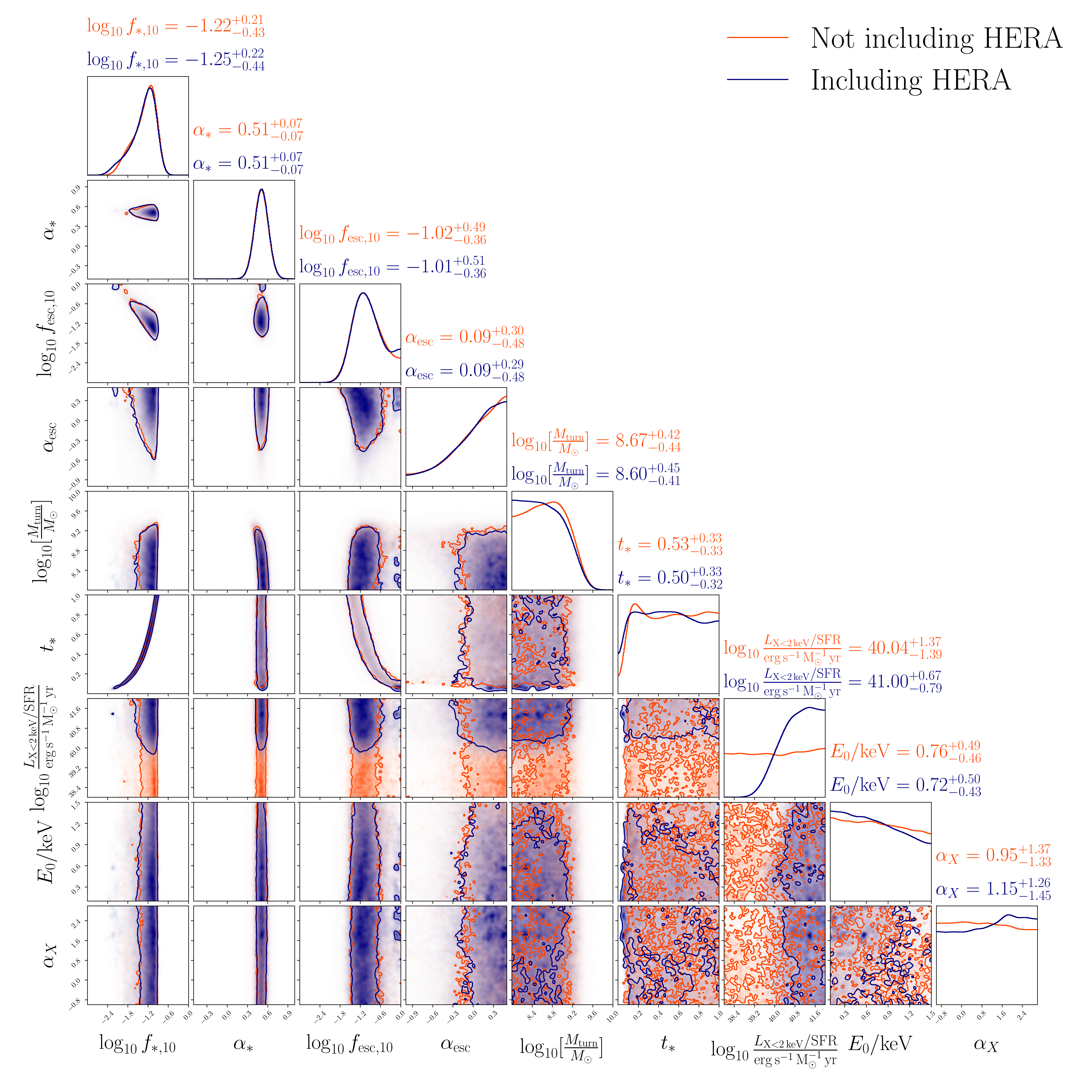}\\
\vspace{-0.25in}
    \caption{Posterior distribution with and without HERA data. Posteriors without HERA data are produces using upper bound on the neutral fraction $\bar{x}_{H\RNum{1}}$ at $z=5.9$, $\tau_e$ - CMB optical depth, and galaxy UV luminosity function at redshift $z$ = 6 - 10 ($M_{\rm UV} > -20$). The most significant effect of adding the HERA data, is ruling out the possibility of low values of $L_{\rm X<2 \, keV}/{\rm SFR}$ and setting the 68\% highest posterior density (HPD) on  $L_{\rm X<2 \, keV}/{\rm SFR} = [10^{40.6}, 10^{42}]$ $\rm erg\, s^{-1}\,M_{\odot}^{-1}\,yr$, which is in good agreement with the results presented at \cite{HERA:2022wmy}, $L_{\rm X<2 \, keV}/{\rm SFR} = [10^{40.4}, 10^{41.7}]$ $\rm erg\, s^{-1}\,M_{\odot}^{-1}\,yr$.}
    \label{fig:mcmc_results}
\end{figure*}

In Fig.~\ref{fig:mcmc_results}, we show 1D and 2D marginalized posterior distributions for
inferences I-A and I-B. This figure is directly comparable to Fig.~6 of Ref.~\cite{HERA:2021noe}.
We find that our results resemble the results of Ref.~\cite{HERA:2021noe} except for 
one detail.
In Ref.~\cite{HERA:2021noe}, we see that there is a sudden decline in the 1D posterior at the two ends
for parameters $\alpha_{\rm X}$ and $E_0$, and at the highest end for 
$t_{\ast}$ and $\Lx$. The reason for this in the case of $\Lx$
was discussed previously, and it is likely due to the unequally-weighted sample points in \texttt{21cmMC},
and we suspect the same is happening here for the other parameters.

\begin{figure*}
    \centering
\includegraphics[width = \textwidth]{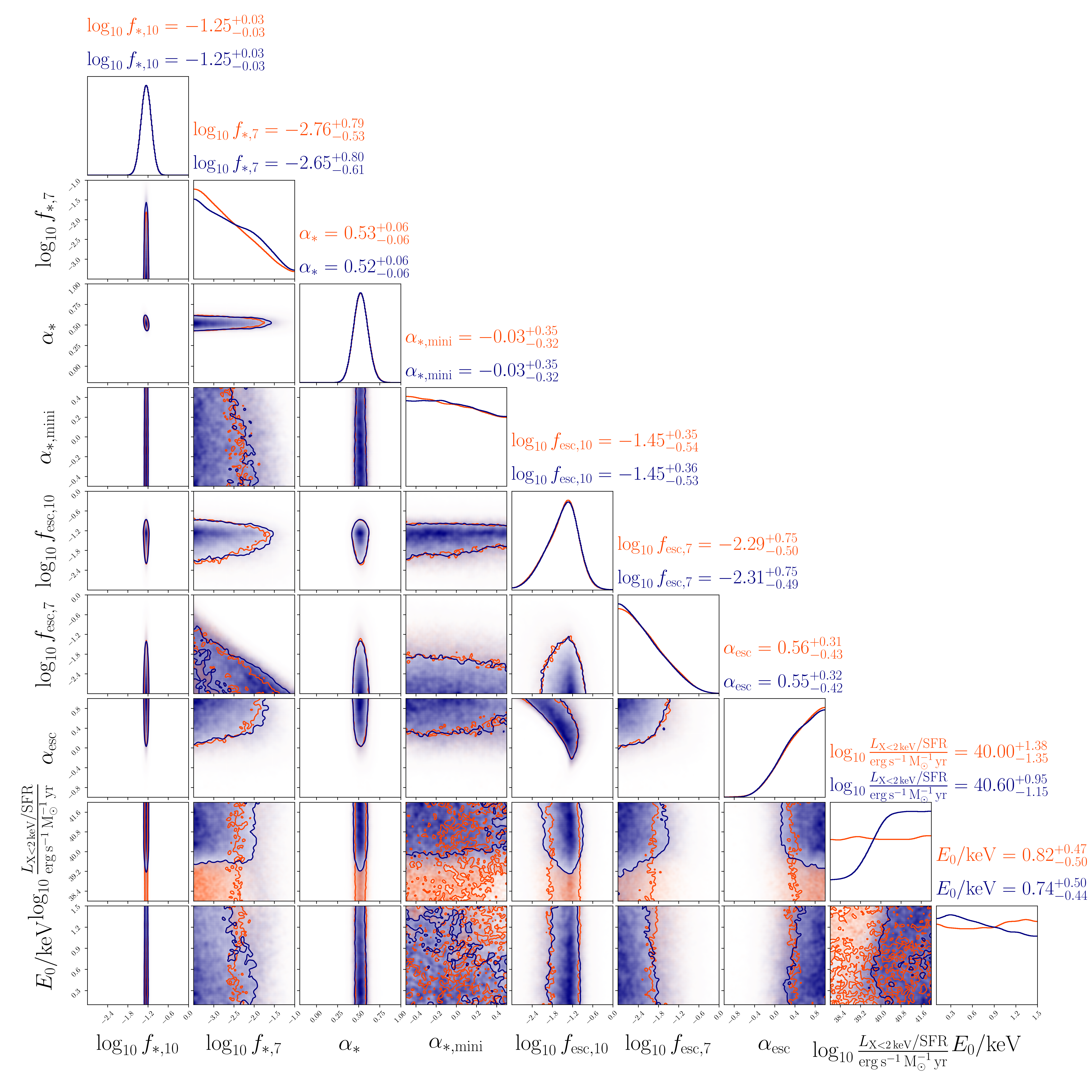}
\vspace{-0.25in}
\caption{Posterior distribution with and without HERA data, including MCGs. The most significant difference from what presented at figure \ref{fig:mcmc_results}, are the relaxed constraints on $L_{\rm X<2 \, keV}/{\rm SFR}$, setting the 68\% highest posterior density (HPD) on  $L_{\rm X<2 \, keV}/{\rm SFR} = [10^{40}, 10^{42}]$ $\rm erg\, s^{-1}\,M_{\odot}^{-1}\,yr$, and the 95\% HPD on $[10^{38.6}, 10^{42}]$ $\rm erg\, s^{-1}\,M_{\odot}^{-1}\,yr$.}
\label{fig:mcmc_results_mini_halos}
\end{figure*}

Fig.~\ref{fig:mcmc_results_mini_halos} is the same as Fig.~\ref{fig:mcmc_results} but for
inferences II-A and II-B. Here also most of the parameters are largely constrained by 
the external data, and HERA data only helps to put a lower bound on $\Lx$. 
Median values of most of the ACG parameters are almost similar to Fig.~\ref{fig:mcmc_results}
except for $f_{{\rm esc},10}$. With MCGs in the scenario, we find that a smaller median value
of $f_{{\rm esc},10}$ is now preferred. This is expected as less radiation is required to escape from 
the ACGs since MCGs also contribute to the ionizing photon budget. 
We also see that $\log_{10}f_{\ast,10}$ here is more tightly constrained than in 
Fig.~\ref{fig:mcmc_results}, although the median values are similar in both figures. 
Considering the MCG parameters, 
we find that the total likelihood places only upper bounds on $\log_{10}f_{\ast,7}$ and 
$\log_{10}f_{{\rm esc},7}$, and almost no statistically significant bound on $\alpha_{\ast,\rm mini}$. 

\subsection{Considering different priors on MCGs parameters}
\begin{figure*}[t!]
    \centering
    \includegraphics[width = \columnwidth]{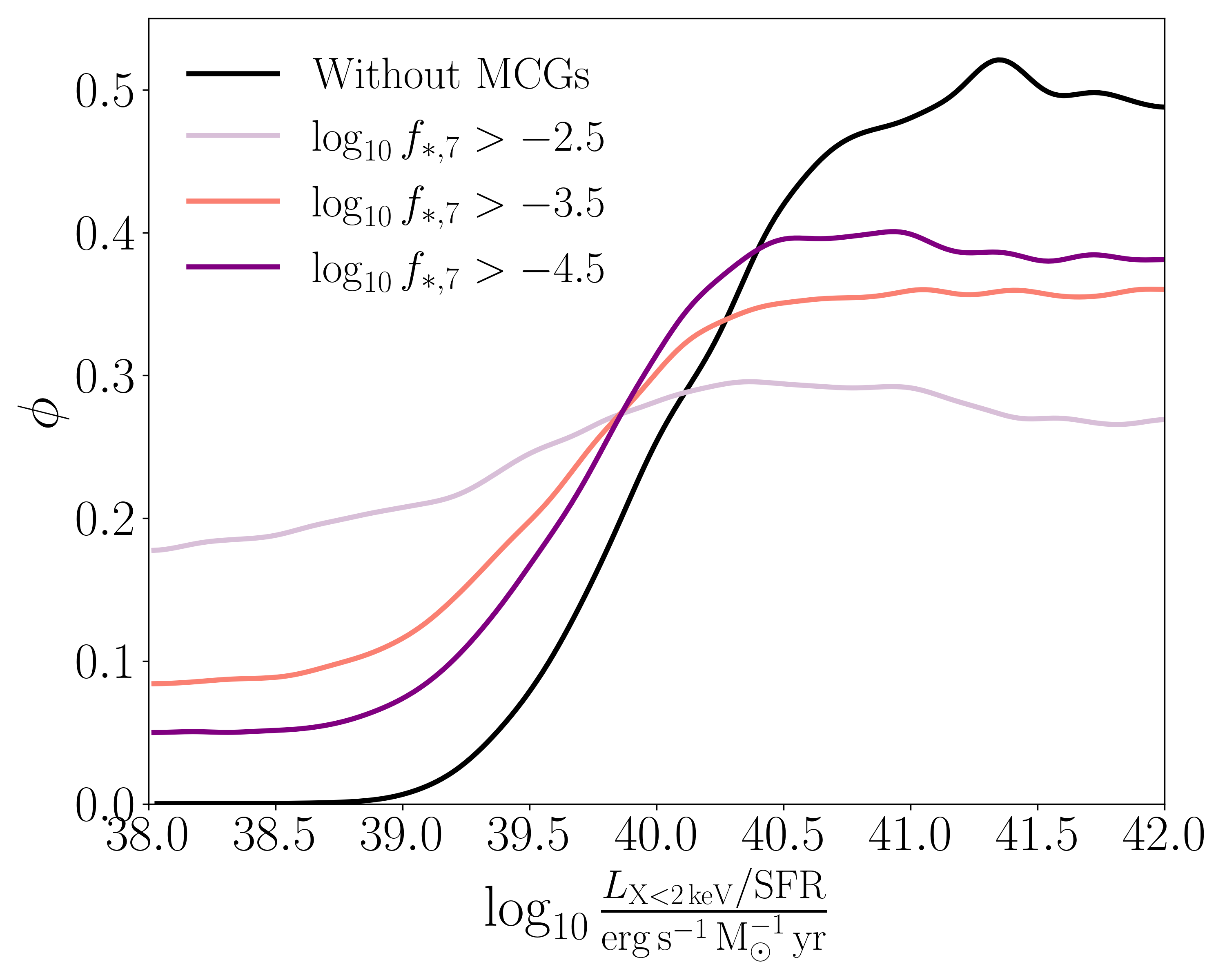}
    \includegraphics[width = \columnwidth]{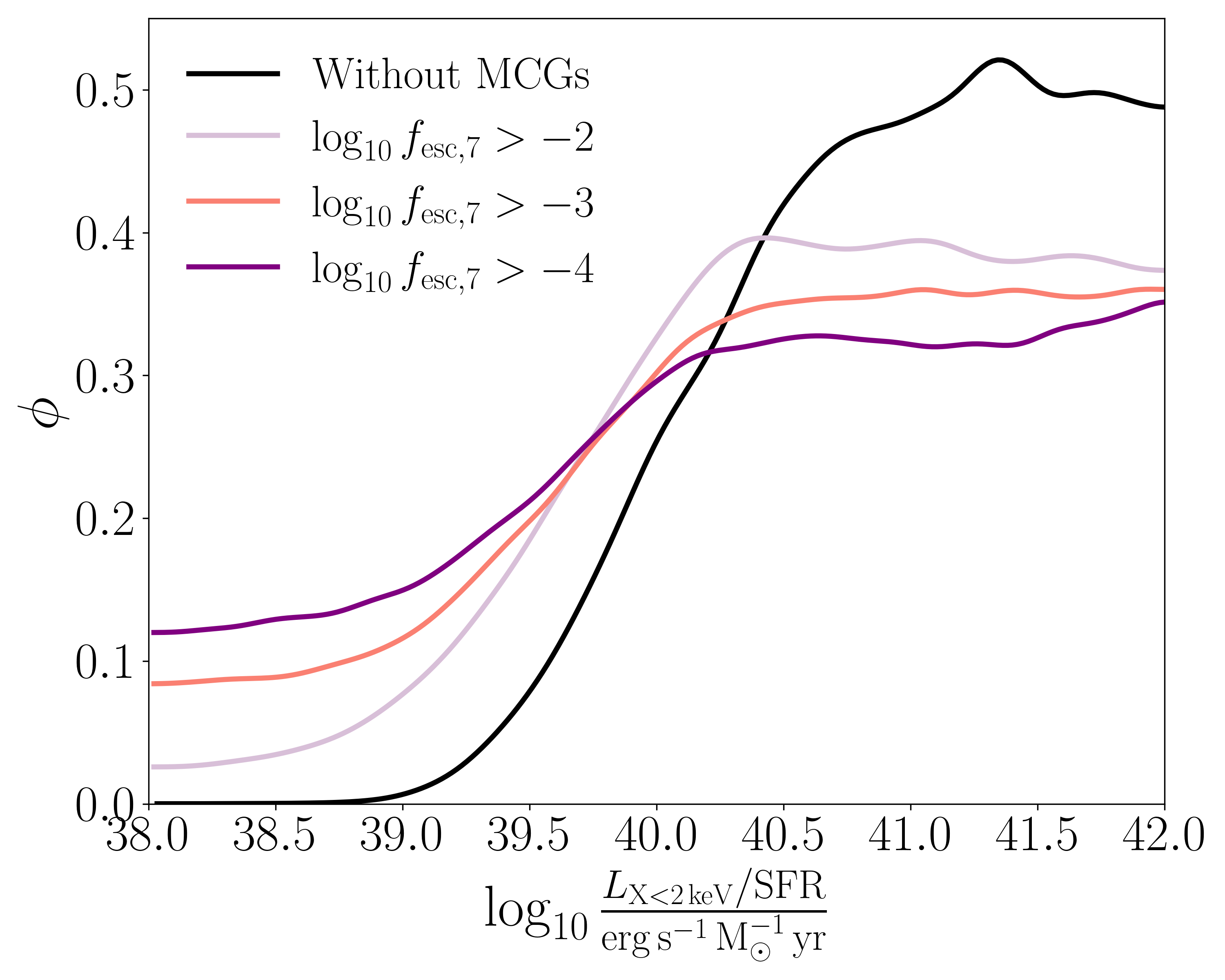}
    \caption{Marginalized posterior distribution of $\Lx$ obtained when including MCGs, while varying   the lower bound of $\log_{10}f_{\ast,7}$ ({\it Left}) or  the lower bound of $\log_{10}f_{{\rm esc},7}$ ({\it Right}). When lower values of $\log_{10}f_{\ast,7}$ are enabled, the posterior distribution tends towards the results of I-B (ACGs only). In case we restrict $\log_{10}f_{\ast,7}$ to a higher range, the $\Lx$ constraints tend to vanish, as less X-ray luminosity is now required to lower the amplitude of the signal at the redshifts measured by HERA. Similarly, when higher values of $\log_{10}f_{{\rm esc},7}$ are forced, the posterior distribution tends towards the results of I-B (ACGs only). If we allow $\log_{10}f_{{\rm esc},7}$ to be smaller, the $\Lx$ constraints tend to relax.}
    \label{fig:f_star7_distribution}
\end{figure*}

As discussed in~\cite{Qin:2020xyh}, it is important to note that the constraints over $\Lx$, achieved when introducing MCGs, are highly dependent on the allowed values of $\log_{10}f_{\ast,7}$ and $\log_{10}f_{{\rm esc},7}$, i.e.\ on the choice of priors. In the above analysis, we did not allow $\log_{10}f_{\ast,7}$ to be smaller than -3.5. When lower values are permitted, the external likelihoods push $\log_{10}f_{\ast,7}$ towards them. The star-formation-rate density (SFRD) of MCGs depends linearly on $f_{\ast,7}$ \cite{Munoz:2021psm}, thus it decreases as well. In turn, that decreases the number of X-ray producing remnants in MCGs, which increases the overall $\Lx$ required so that the power spectrum amplitude remains below the HERA upper bounds. The resulting constraints are very similar to the ones produced when taking only ACGs into account, as shown in Fig.~\ref{fig:f_star7_distribution} (left panel). The opposite effect occurs when increasing the lower bound on $\log_{10}f_{\ast,7}$. Higher values of $\log_{10}f_{\ast,7}$, i.e.\ higher MCGs SFRD, yield more X-ray producing sources, thus decreasing the required $\Lx$ for the signal to be consistent with the current HERA data. This behaviour is illustrated as well in Fig.~\ref{fig:f_star7_distribution} (left panel).

Similar behaviour is seen when the lower bounds on $\log_{10}f_{{\rm esc},7}$ are varied. Our fiducial value is $\log_{10}f_{{\rm esc},7}$ $> -3$.
When $\log_{10}f_{{\rm esc},7}$ is forced to be larger, ionization occurs earlier, thus increasing the optical depth ($\tau$ is calculated as an integral, where the integrand is proportional to $1-x_{HI}$). In order to fit for the optical depth constraint from Ref.~\cite{Planck:2018vyg}, $\log_{10}f_{\ast,7}$ must be smaller, and so the $\Lx$ posterior tends towards the one achieved in analysis I-B, as discussed above. In the opposite scenario, we enable lower values of $\log_{10}f_{{\rm esc},7}$, and two related processes take place. Since $\log_{10}f_{{\rm esc},7}$ is smaller, ionization occurs later, such that the optical depth is smaller as well. In addition, the neutral fraction at $z=5.9$, also increases when $\log_{10}f_{{\rm esc},7}$ is lowered. The constraints on these two observables force a higher value of $\log_{10}f_{\ast,7}$, and as discussed above this enhances the effect of including MCGs is the analysis. The $\Lx$ posteriors in the different cases are shown in Fig.~\ref{fig:f_star7_distribution} (right panel). The bottom line is that the above analysis is prior dependent, and caution is warranted when reaching conclusions for models with MCGs.

\section{Conclusions}\label{section_6}

In this work, we have introduced an ANN-based emulator of the 21-cm signal which was trained on 
data produced by the \texttt{21cmFAST} semi-numerical code. We found this emulator to be  accurate enough
 to predict the 21-cm signal over a range of redshifts $z$ and wave numbers $k$. 

We then used this
emulator in an MCMC pipeline to reproduce the parameter constraints presented in 
Refs.~\cite{HERA:2021noe,HERA:2022wmy}, based on the HERA phase-I upper limits on the 21-cm power spectrum and 
the external likelihoods containing information on the high-redshift UV luminosity function, the IGM neutral
fraction and the reionization optical depth. Here, the underlying astrophysical model 
assumption is that the atomically cooled galaxies (ACGs) (which host the 
PopII stars) sparked the cosmic dawn and are responsible for the subsequent 
heating and reionization of the IGM. Our emulator-based MCMC pipeline seems to
produce similar results as presented in  Ref.~\cite{HERA:2021noe}, which validates our pipeline. 
One of the most important results of this analysis, based on the posterior of 
the parameter $\Lx$, can be stated as 
follows. If PopII stars (or ACGs) dominate the X-ray heating and reionization of the IGM, then we expect
the high redshift ($z>6$) galaxies to be more X-ray luminous 
(with $\Lx \gtrsim 10^{40} \Lu$)
than their present-day counterparts (for which $\Lx \approx 10^{39} \Lu$)
~\cite{Mineo:2011id,Lehmer:2010en}. 

Our primary goal in this
work was to check the validity of the above result when we include molecular-cooled galaxies (MCGs) in the analysis. 
A number of recent simulations show that MCGs, 
which predominantly host PopIII stars, also contribute to the total photon budget during cosmic dawn and reionization. 
Ignoring MCGs in any analysis could therefore lead to an overestimation of
the X-ray and ionizing efficiencies of the atomic cooling galaxies (ACGs). 
The same concern has also been mentioned in Ref.~\cite{HERA:2022wmy}.

We therefore trained our emulator with the results obtained after including PopIII stars in the simulations,
and finally ran an MCMC analysis that fits for additional parameters corresponding to the PopIII stars.
The most interesting result that emerges out of the final posterior distributions we obtained 
is that including both ACGs and MCGs in the simulations relaxes the preference for the high X-ray
luminosity of the high redshift sources as we discussed in the previous paragraph, 
and now values as low as $\Lx \lesssim 10^{39} \Lu$ are still allowed by the HERA power spectrum
data. 
This is due to the fact that MCGs contribute to the X-ray heating and therefore the ACGs 
do not need to be as X-ray efficient. This indicates that the X-ray luminosity of the high redshift sources may not be very different from their low redshift counterparts.
It is important to note, that although we do see a small decline in the posterior distribution at the lower $\Lx$ regime when including MCGs (as in Figs.~\ref{fig:Lx_marg_all} and \ref{fig:Lx_marginalized}) we cannot determine confidently that higher values are preferred. This is because the chosen likelihood function (see Eq.~\eqref{eq:final_marg_likelihood}) gives higher probability for power spectra well below the HERA upper bounds. In fact, as we see in Fig.~\ref{fig:Lx=38_mini}, the spectra of this low X-ray signal is just below the HERA bound, making it a valid scenario. Contrary to that, the same figure shows that when MCGs are not included, the spectra obtained with large $\Lx$ is well above the upper bound, As mentioned above, this result depends on the choice of priors, and specifically on the allowed ranges of $\log_{10}f_{\ast,7}$ and $\log_{10}f_{\ast,7}$, and caution has to be taken when applying this conclusions.

\begin{figure}[h]
    \centering
    \includegraphics[width = \columnwidth]{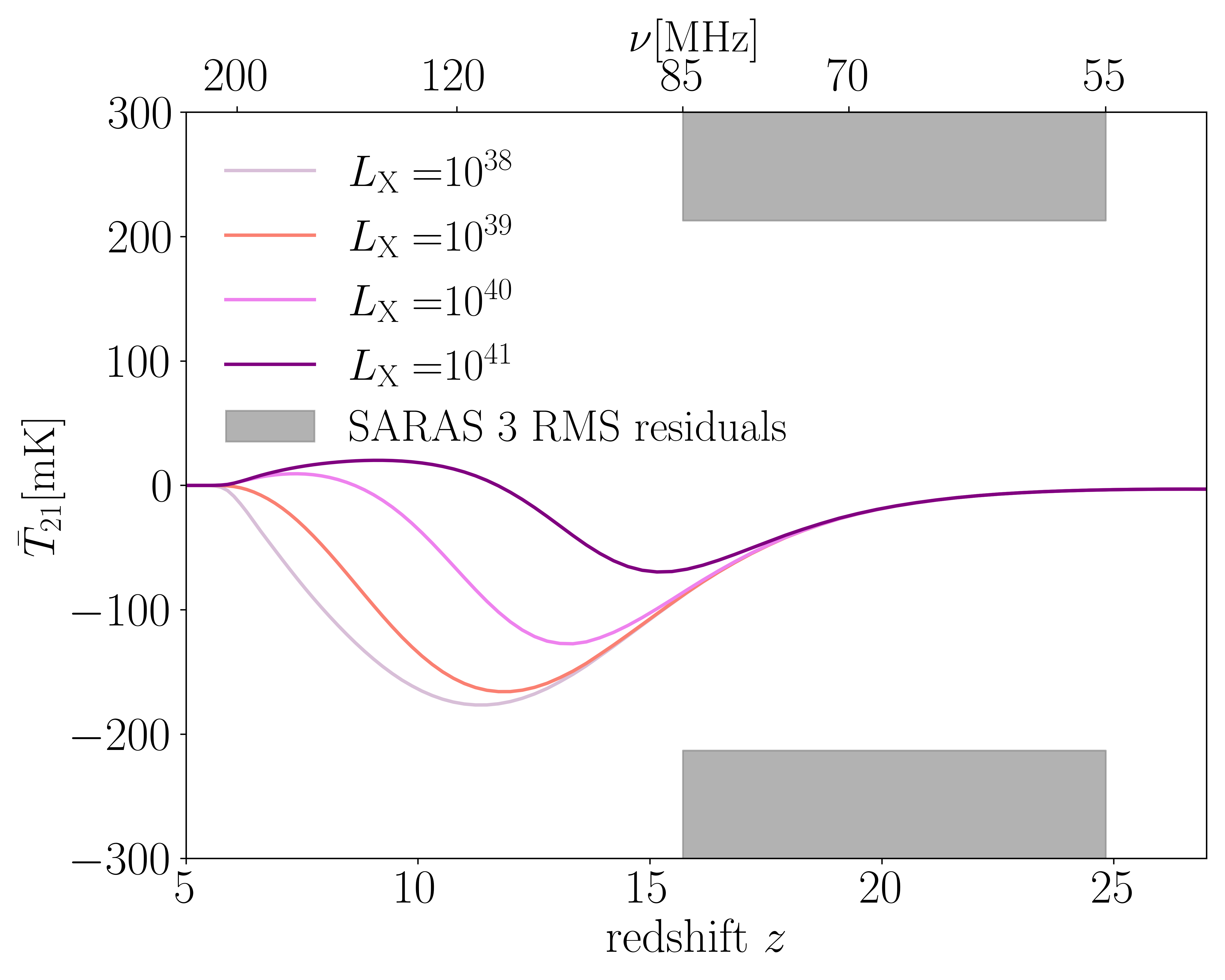}
    \caption{Comparison of the 21-cm global signal obtained with median values from the posterior distribution from the inference which accounts for MCGs, and
    different $\Lx$ values. These values are compared to the SARAS 3 brightness temperature bounds,  demonstrating that introducing MCGs while enabling lower values of $\Lx$ does not create a conflict with earlier cosmological measurements. }
    \label{fig:Tb_compare}
\end{figure}

Before closing, we check the  validity of our results by comparing them with the 21-cm global signal 
$\langle \delta T_{21} \rangle$ measured by 
the Shaped Antenna Measurement of the Background Radio Spectrum 3 (SARAS 3)~\cite{Singh:2021mxo}
experiment in the frequency band $55 \rm MHz$ to $85 \rm MHz$. Note that SARAS 3 did not yield a statistically significant detection of $\langle \delta T_{21} \rangle$, rather they provide 
RMS residuals of amplitude $213 \rm mK$ which can be considered as upper and lower bounds of the signal. 
Fig.~\ref{fig:Tb_compare} shows that our results do not contradict the 
SARAS 3 results. Note that existing global signal 
\cite{Bowman:2010qdu,Monsalve:2017mli,Monsalve:2018fno,Monsalve:2019baw,Singh:2021mxo}
measurements do not have much constraining power and hence are not included in our analysis. However, with more precise experiments in the future, global signal
measurements will achieve more constraining power. In that case, our pipeline can be easily extended to 
emulate the global signals \cite{Bye:2021ngm, Cohen:2019vck, Bevins:2021eah}. We shall explore
this possibility in future work. 

Note that in our analysis, we have assumed that the X-ray luminosities of both the 
Pop\RNum{2} and Pop\RNum{3} stars are same, i.e. 
 $L^{\rm (II)}_{\rm X<2 \, keV}/{\rm SFR}$ = $L^{\rm (III)}_{\rm X<2 \, keV}/{\rm SFR}$ =
 $\Lx$. We call $\Lx$ as the effective X-ray luminosity. 
 This choice is made just to reduce the total number of parameters. 
 However, this choice affects the prior volume in a non-trivial manner. 
 Making $L^{\rm (II)}_{\rm X<2 \, keV}/{\rm SFR}$ = 
 $L^{\rm (III)}_{\rm X<2 \, keV}/{\rm SFR}$ automatically reduces the effective value of 
 $\Lx$. This would not be the case if $L^{\rm (III)}_{\rm X<2 \, keV}/{\rm SFR}$ is made
 to vary independently. In this case, if the likelihood had preferred a lower value 
 $L^{\rm (III)}_{\rm X<2 \, keV}/{\rm SFR} << L^{\rm (II)}_{\rm X<2 \, keV}/{\rm SFR}$, 
 we would have obtained the same conclusion as in Ref~\cite{HERA:2021noe}. 
 However, this is unlikely as
 the theoretical modelling indicates that the low-metallicity stars, like Pop\RNum{3},
 are expected to exhibit high $\Lx$ \cite{Fragos:2012vf}, and we anticipate 
 $L^{\rm (III)}_{\rm X<2 \, keV}/{\rm SFR} \gtrsim L^{\rm (II)}_{\rm X<2 \, keV}/{\rm SFR}$. Therefore, we expect that our conclusions
 will still be valid under the assumption 
 $L^{\rm (II)}_{\rm X<2 \, keV}/{\rm SFR}$ = $L^{\rm (III)}_{\rm X<2 \, keV}/{\rm SFR}$ =
 $\Lx$.

Future study will also try to address the limitations of the emulator used in this work, most significant of which is
the inability to predict the flat or zero signals reliably. We have proposed a workaround 
considering the present scenario. However, with a more precise measurement of the power spectrum, 
we will need to predict signals of all kinds with better precision. Below, we discuss a few possible approaches 
to improving our emulation pipeline: (i) Using Generative Adversarial Networks (GANs) 
\cite{Goodfellow:2014upx}, which implement some unsupervised learning methods to train models.
(ii) Training the ANN on the likelihood function instead of the power spectrum.
In that case, the ANN does not have to learn the complicated features of the signal.
Rather, it learns to make decisions as to how acceptable the signal is based on the likelihood. (iii) Since in future experiments we expect to have more redshift bands, we can treat the power-spectra observations as two dimensional data, where one dimension is the redshift, and the other is the wave number. We therefore can exploit the developing research in the field of Convolutional Neural Networks (CNNs) \cite{ciresan2011flexibles, Szegedy:2014nrf}, which uses for machine learning algorithms where the data has more than one dimension,
for our goal. A different approach to the computational problem of producing samples of our model in a reasonable time, is using analytic codes like \texttt{Zeus21}~\cite{Munoz:2023kkg}. This code provides good approximations for all of \texttt{21cmFAST} (ACG-only) outputs, in $\mathcal{O}(1-10)\,{\rm sec}$.

The conclusions of our work may have a number of consequences which we note below: 
(i) MCGs spark cosmic dawn and consequently, the Ly-$\alpha$ coupling starts early.
This allows the spin temperature to catch up with the kinetic temperature at a higher redshift compared to the
scenario where only ACGs exist. Now, a combination of less X-ray luminous 
($\Lx \lesssim 10^{40} \Lu$) MCG and ACGs would make the absorption signal deeper and 
extended over several redshifts 
(compared to more luminous $L_{\rm X<2 \, keV}/{\rm SFR} > 10^{40} \Lu$ sources). This bodes well for
global signal experiments as deeper and more extended absorption signals are easier to detect. 
(ii) Since the amplitude of the 21-cm power spectrum increases as we decrease $\Lx$, this could also be fortunate for experiments like HERA and SKA that target 21-cm fluctuations. 
(iii) Low X-ray efficiency of the early sources provides room for  other heating mechanisms
to be effective, such as Ly-$\alpha$ heating due to  resonant 
scattering of the Ly-$\alpha$ photons with the atoms in IGM~\cite{Reis:2021nqf,Sarkar:2022dvl}. 

It is crucial to take caution when interpreting results related to 
the modelling of the 21cm signal at high redshifts, as these outcomes 
heavily hinge on the intricate modelling of various processes. Notably, different codes employing similar source populations but employing distinct algorithms may yield disparate results.  \texttt{21cmFAST}
is one such model that uses a specific algorithm to model the cosmic dawn and reionization.
It would be worth checking the consistency of the results presented here with other similar
codes.

 Finally, we emphasize that the emulator-based analysis pipeline developed in this work is flexible, fast and does not require a prohibitive number of simulations. We therefore plan to employ it to place 21-cm bounds on models of dark matter~\cite{Kovetz:2018zan,Kovetz:2018zes,Flitter:2022pzf}, primordial magnetic fields~\cite{Adi:2023qdf} and various other extensions to $\Lambda$CDM.

\vspace{-0.1in}
\begin{acknowledgements}

\vspace{-0.05in}

We thank Rennan Barkana, Daniela Breitman, Anastasia Fialkov, Jordan Flitter and Caner Unal for useful discussions. We are grateful to Andrei Mesinger and Julian Mu\~noz for insightful comments on the manuscript. We also acknowledge the efforts of the entire {\tt 21cmFAST} team in producing a state-of-the-art 21-cm public simulation code. HL is supported by a Dkalim excellence M.Sc.\ scholarship from Ben-Gurion University. EDK is supported by an Azrieli Foundation faculty fellowship.

\end{acknowledgements}

\clearpage

\appendix
\section{Emulators for $\tau_e$ and $x_{H\RNum{1}}$}\label{apndx_1}
The emulators for $\tau_e$ and $x_{\rm H\RNum{1}}$ are built on the same structure 
described in Section~\ref{section_3}. 
The data set for each observable contain $\sim$ 20000 simulations where the parameters are sampled using 
the LH \cite{McKay1979} sampler. The whole data is randomly divided into 
training (85\%), validation (10\%) and testing (5\%) sets. The astrophysical parameters are normalized to fit in the
range $[-1,1]$ for convenience. The NN here is composed of 3 layers of sizes 256, 512, 512 respectively. We have used 
the ReLU \cite{4082265} activation function, and we have checked that it performs really well. 
As done for the emulators described in section \ref{section_3}, we added a batch Normalization layer \cite{Ioffe:2015ovl} after each activation.
For the loss function, we rely on the standard L2 norm, defined as 
\begin{equation}
    \mathcal{L}(y_{true},y_{pred}(\theta)) = \frac{1}{m}\sum_{i=1}^m (y_{true} - y_{pred}(\theta))^2 
    \,,
\end{equation} 
where $m$ is the number of samples in a batch which is taken to be 128. 
Here, the Adam \cite{Kingma:2014vow} optimizer is used in the training with an initial learning rate of 0.01. 
Depending on the performance of the ANN on the validation set, 
we even reduce the learning rate by a factor of two. In the end, we restore the best weights.

\begin{figure}[h]
    \centering
    \includegraphics[width = \columnwidth]{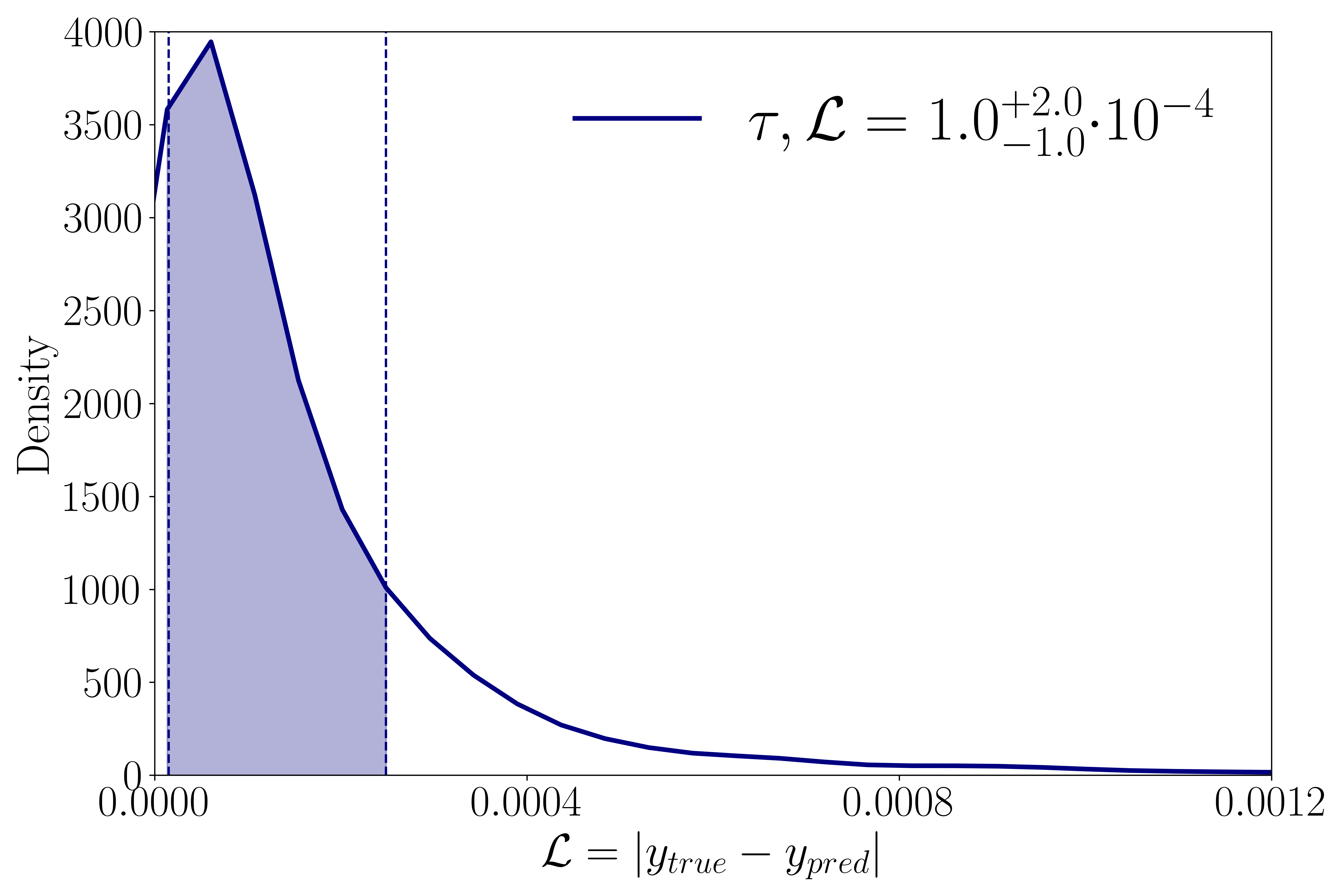}
    \caption{Loss statistics for the $\tau$ emulator based on testing sets having $\sim$ 1000 samples each.
    The dashed lines represent the 16th and 84th percentiles respectively, and their exact value is written along
    with the median value in the figure legend. Note that the X-axis here is not normalized with the true values.
    }
    \label{fig:tau_results}
\end{figure}

\begin{figure}[h]
    \centering
    \includegraphics[width = \columnwidth]{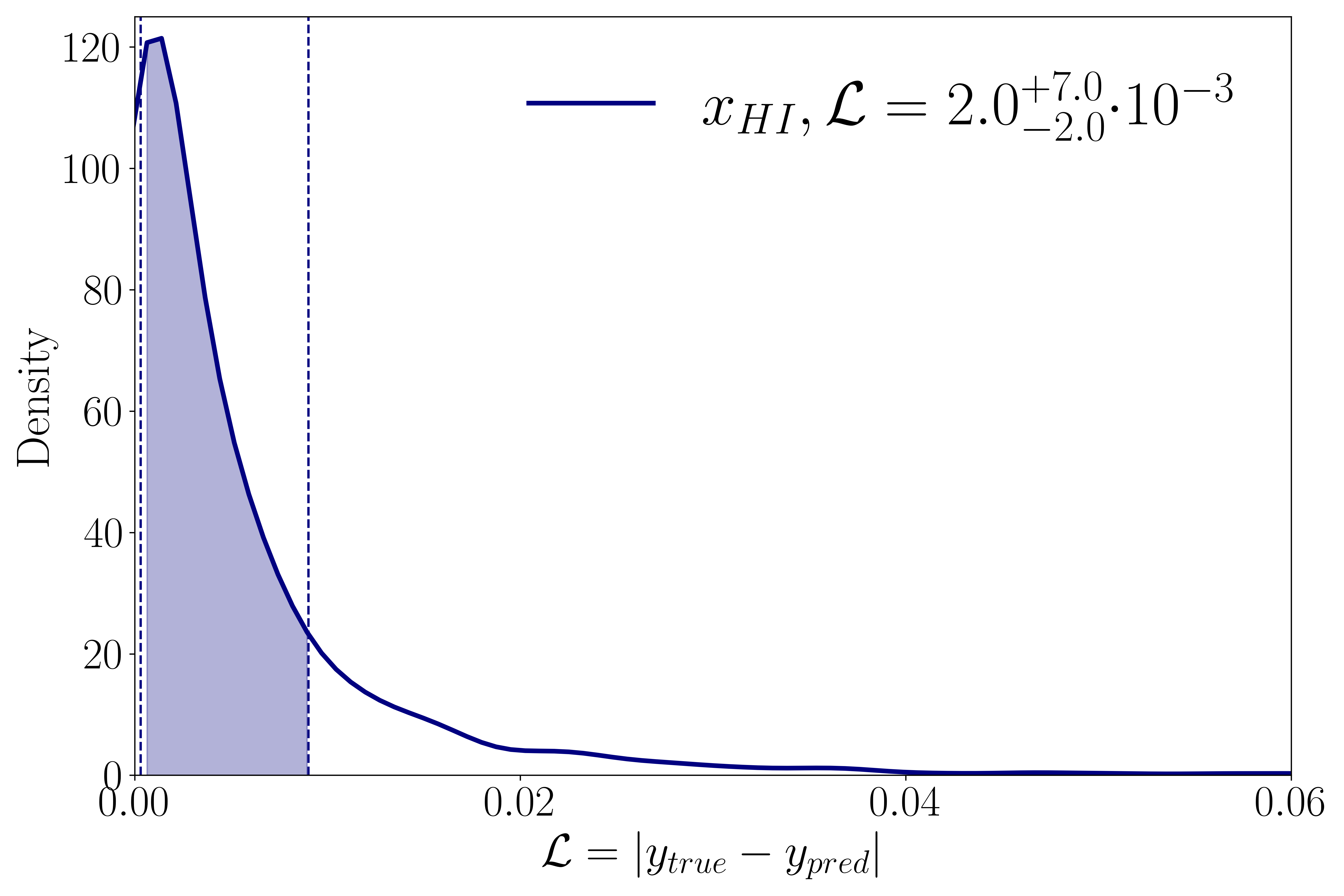}
    \caption{
    Loss statistics for the $x_{\rm HI}$ emulator based on testing sets having $\sim$ 1000 samples each.
    The dashed lines represent the 16th and 84th percentiles respectively, and their exact value is written along
    with the median value in the figure legend. Note that the X-axis here is not normalized with the true values.
    }
    \label{fig:xH_results}
\end{figure}

As can be seen from Figures~\ref{fig:tau_results} and \ref{fig:xH_results}, the emulators perform really well. 
Even the rare errors are not significant. This convinces us that the constraints obtained from these observables 
will not suffer from emulation errors.

\section{Emulator with MCGs}\label{apndx_b}

\begin{figure}[h]
    \centering
    \includegraphics[width = \columnwidth]{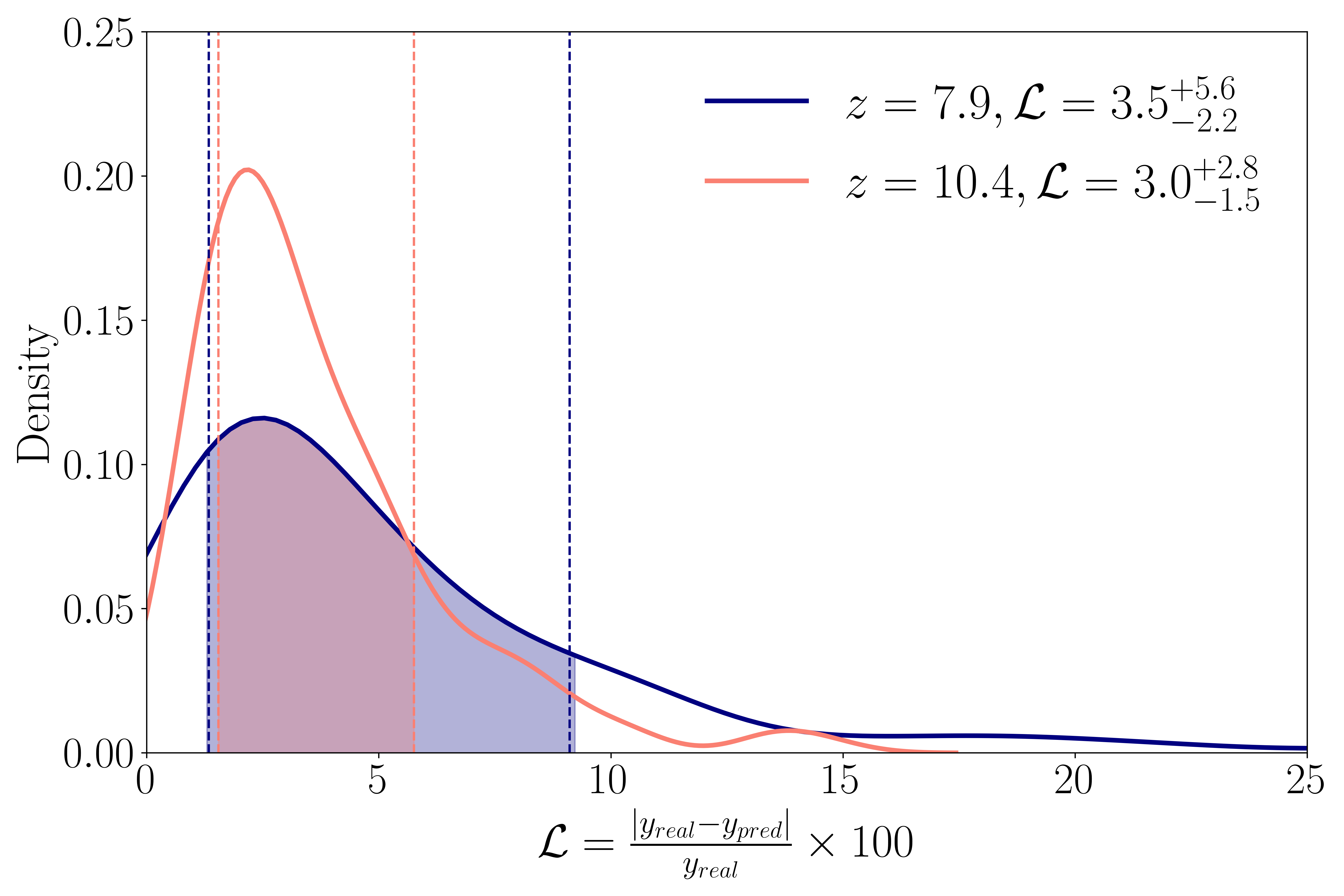}
    \caption{
    Loss statistics for the power spectrum emulator, where MCGs are included in the simulations.  
    Results are based on testing sets having $\sim$ 250 samples in each.
    The dashed lines represent the 16th and 84th percentiles respectively, and their exact value is written along
    with the median value in the figure legend.
    }
    \label{fig:emulator_mini_retrained_testing}
\end{figure}

Here we present the performance of our emulator where the simulations used to train it include both ACGs and MCGs. 
The pipeline and procedures used here are the same as described in Section~\ref{section_3}, 
and we do not repeat these. We present the emulation results in Figure~\ref{fig:emulator_mini_retrained_testing}. 
We find that the results are almost similar to what we found for the emulator where only ACGs are present.

\section{Classifiers}\label{apndx_c}

In this work, we are using a total of four different signal classifiers for the power spectrum. 
One for each redshift $(7.9, 10.4)$, and 
for each redshift, there are two sub-classifiers for emulation with (i) only ACGs and (ii) ACGs+MCGs.
For each classification, we have used $\sim 10000$ signals and divided those 
into training (85\%) validation (10\%) and testing (5\%) sets. 
We have also normalized the astrophysical parameters to fit the range $[-1,1]$ for faster and more efficient learning.

The NN for all the classifiers is composed of three layers of sizes 128, 512, and 256 respectively.
For activation, we have used ReLU \cite{4082265} followed by a batch Normalization layer 
\cite{Ioffe:2015ovl} which standardizes the input for the next layer,
and a Dropout \cite{hinton2012improving} layer, which randomly turns off $10\%$ of the neurons 
in the previous layer in order to prevent over-fitting. The end layer contains one neuron that
uses a `sigmoid' activation function. This is chosen in order to have the classifier output in the range 
$\in[0,1]$. Here, an output smaller than $0.5$ is classified as class 0 and  $\geq 0.5$ is 
classified as class 1 (as defined in Section~\ref{sec:classification}). 
We use the Binary cross Entropy loss function, which is the standard choice for 
classifying a data set into two classes, defined as,
\begin{equation}
    \mathcal{L} = -\frac{1}{m} \sum_{i=1}^m y_i \log (P(y_i)) + (1-y_i)\log (1-P(y_i))\,,
\end{equation}
where $y_i$ are the labels, $P(y_i)$ is the classifier probability prediction for this label, 
and $m$ is the number of samples in a batch which we have chosen to be $512$.
We have used the Adam \cite{Kingma:2014vow} optimizer with an initial learning rate of 0.01, which is
reduced by a factor of two based on the performance. We store the best weights at the end. 
We found that the accuracy is $\gtrsim 97\%$ for all the classifiers, and we can confidently use this in the
MCMC process.

\end{document}